\documentclass[aps,prb,twocolumn,footinbib,superscriptaddress,longbibliography]{revtex4-2}
\usepackage{graphicx}
\usepackage{float}
\usepackage{bm}
\usepackage{amsmath,amssymb}
\usepackage[cal=boondoxo]{mathalfa}
\allowdisplaybreaks
\usepackage[bookmarksnumbered,bookmarksopen]{hyperref}
\hypersetup{
   colorlinks=true,       
   citecolor=blue,        
}

\usepackage{array}
\newcolumntype {s}[1]{@{\hspace{#1}}} 
\newcolumntype {R}{>{$}r<{$}}         
\newcolumntype {C}{>{$}c<{$}}         
\newcolumntype {L}{>{$}l<{$}}         
\newcolumntype {f}{@{\extracolsep\fill}}  

\newcommand* {\vek}[1]{{\ensuremath{\bm{\mathrm{#1}}}}}
\newcommand* {\kk}{\vek{k}}
\newcommand* {\rr}{\vek{r}}
\newcommand* {\bra}[1]{\ensuremath{\langle {#1} |}}
\newcommand* {\ket}[1]{\ensuremath{| {#1} \rangle}}
\newcommand* {\braket}[1]{\ensuremath{\langle {#1} \rangle}}
\newcommand* {\ee}{\ensuremath{\mathrm{e}}}

\usepackage{color}

\graphicspath{{.}{./EPS/}}

\begin{document}

\title{Spherical topological-insulator nanoparticles: Quantum size
effects and optical transitions}

\author{L. Gioia}
\affiliation{Perimeter Institute, 31 Caroline Street North, Waterloo,
Ontario, Canada N2L 2Y5}
\affiliation{Department of Physics and Astronomy, University of
Waterloo, Waterloo, Ontario, Canada N2L 3G1}

\author{M.~G. Christie}
\affiliation{School of Chemical and Physical Sciences and MacDiarmid
Institute for Advanced Materials and Nanotechnology, Victoria
University of Wellington, PO Box 600, Wellington 6140, New Zealand}

\author{U. Z\"ulicke}
\email{uli.zuelicke@vuw.ac.nz}
\affiliation{School of Chemical and Physical Sciences and MacDiarmid
Institute for Advanced Materials and Nanotechnology, Victoria
University of Wellington, PO Box 600, Wellington 6140, New Zealand}
\affiliation{Department of Physics, University of Basel,
Klingelbergstrasse 82, CH-4056 Basel, Switzerland}

\author{M. Governale}
\affiliation{School of Chemical and Physical Sciences and MacDiarmid
Institute for Advanced Materials and Nanotechnology, Victoria
University of Wellington, PO Box 600, Wellington 6140, New Zealand}

\author{A.~J. Sneyd}
\affiliation{School of Chemical and Physical Sciences and MacDiarmid
Institute for Advanced Materials and Nanotechnology, Victoria
University of Wellington, PO Box 600, Wellington 6140, New Zealand}

\date{\today}

\begin{abstract}

We have investigated the interplay between band inversion and size
quantization in spherically shaped nanoparticles made from
topological-insulator (TI) materials. A general theoretical framework
is developed based on a versatile continuum-model description of the
TI bulk band structure and the assumption of a hard-wall mass
confinement. Analytical results are obtained for the wave functions of
single-electron energy eigenstates and the matrix elements for optical
transitions between them. As expected from spherical symmetry,
quantized levels in TI nanoparticles can be labeled by quantum numbers
$j$ and $m=-j, -j+1, \dots, j$ for total angular momentum and its
projection on an arbitrary axis. The fact that TIs are narrow-gap
materials, where the charge-carrier dynamics is described by a type of
two-flavor Dirac model, requires $j$ to assume half-integer values and
also causes a doubling of energy-level degeneracy where two different
classes of states are distinguished by being parity eigenstates with
eigenvalues $(-1)^{j\mp 1/2}$. The existence of energy eigenstates
having the same $j$ but opposite parity enables optical transitions
where $j$ is conserved, in addition to those adhering to the familiar
selection rule where $j$ changes by $\pm 1$. All optical transitions
satisfy the usual selection rule $\Delta m = 0, \pm 1$. We treat
intra- and inter-band optical transitions on the same footing and
establish ways for observing unusual quantum-size effects in TI
nanoparticles, including oscillatory dependences of the band gap and
of transition amplitudes on the nanoparticle radius. Our theory also
provides a unified perspective on multi-band models for charge
carriers in semiconductors and Dirac fermions from elementary-particle
physics.

\end{abstract}

\maketitle

\section{Introduction and synopsis of main results}
\label{sec:intro}

Synthesis and experimental investigation of semiconductor nanocrystals
have been pursued with great effort to elucidate how electronic and
optical properties of such low-dimensional systems are affected by
size-quantization effects~\cite{yof93,efr00}. Recent advances in the
fabrication of nanoparticles made from topological-insulator (TI)
materials~\cite{cho12,ker13a,var14,jia15,var15,cla19,rid19} have
created the opportunity to explore the interplay of topological
properties and quantum confinement~\cite{hon14}. In particular, TI
materials are known to host surface states that are robust to
perturbations and have energies within the bulk band gap~\cite{has10,
has11,has15}. The presence of these surface states is envisioned to
enable intriguing applications for TI-based
optoelectronics~\cite{pol17}, spintronics~\cite{mci12,tia17}, and
thermoelectrics~\cite{xu17}. As device fabrication often involves
nanostructuring of the TI material, it is necessary to understand the
evolution of surface-state properties as the TI nanomaterial's size is
reduced~\cite{zho08,lin09,liu10a,lu10,kot17,gio18}. Our theoretical
description for the electronic and optical properties of spherical TI
nanoparticles sheds new light on the importance and implications of
quantum-size effects in these, and related, nanostructures.

\begin{figure*}[t]
\includegraphics[width=0.323\textwidth]{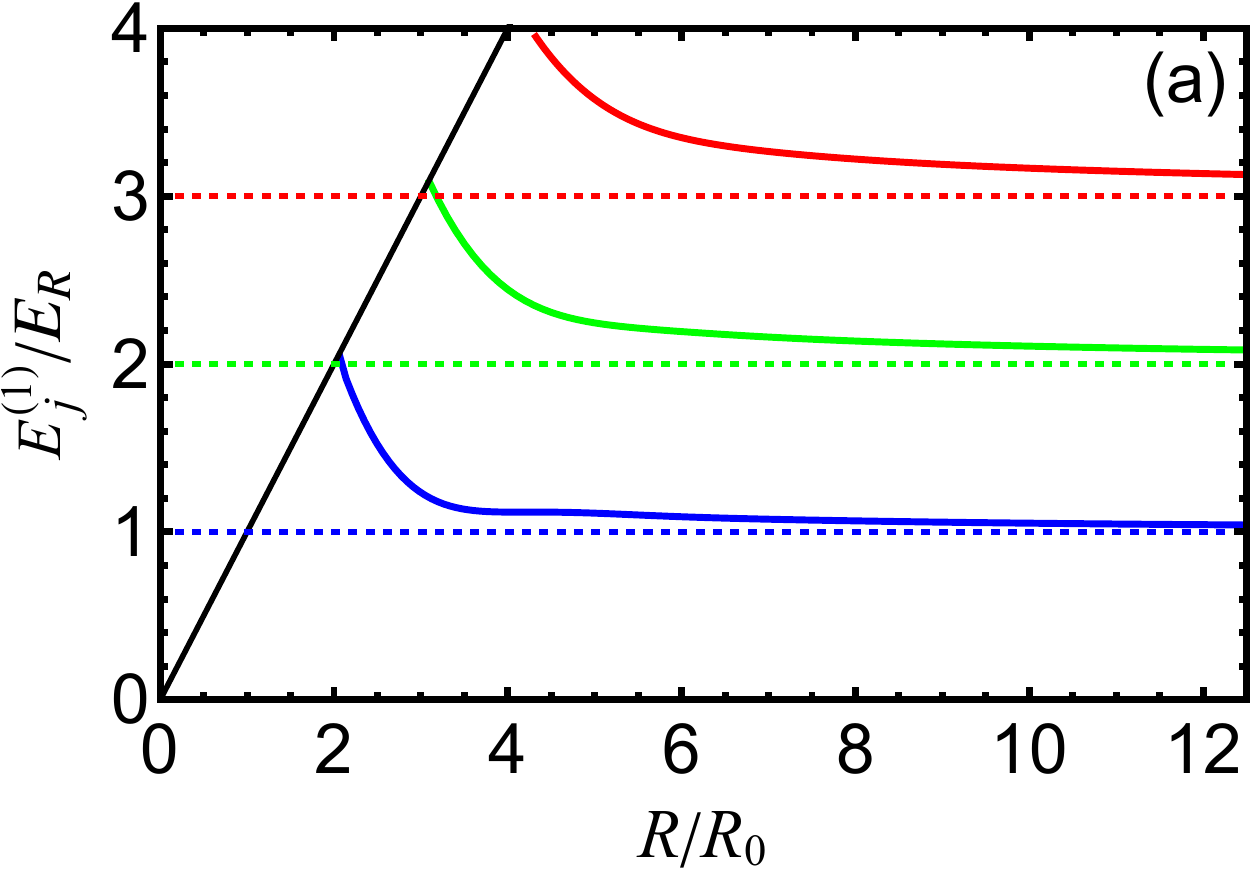}\hfill
\includegraphics[width=0.332\textwidth]{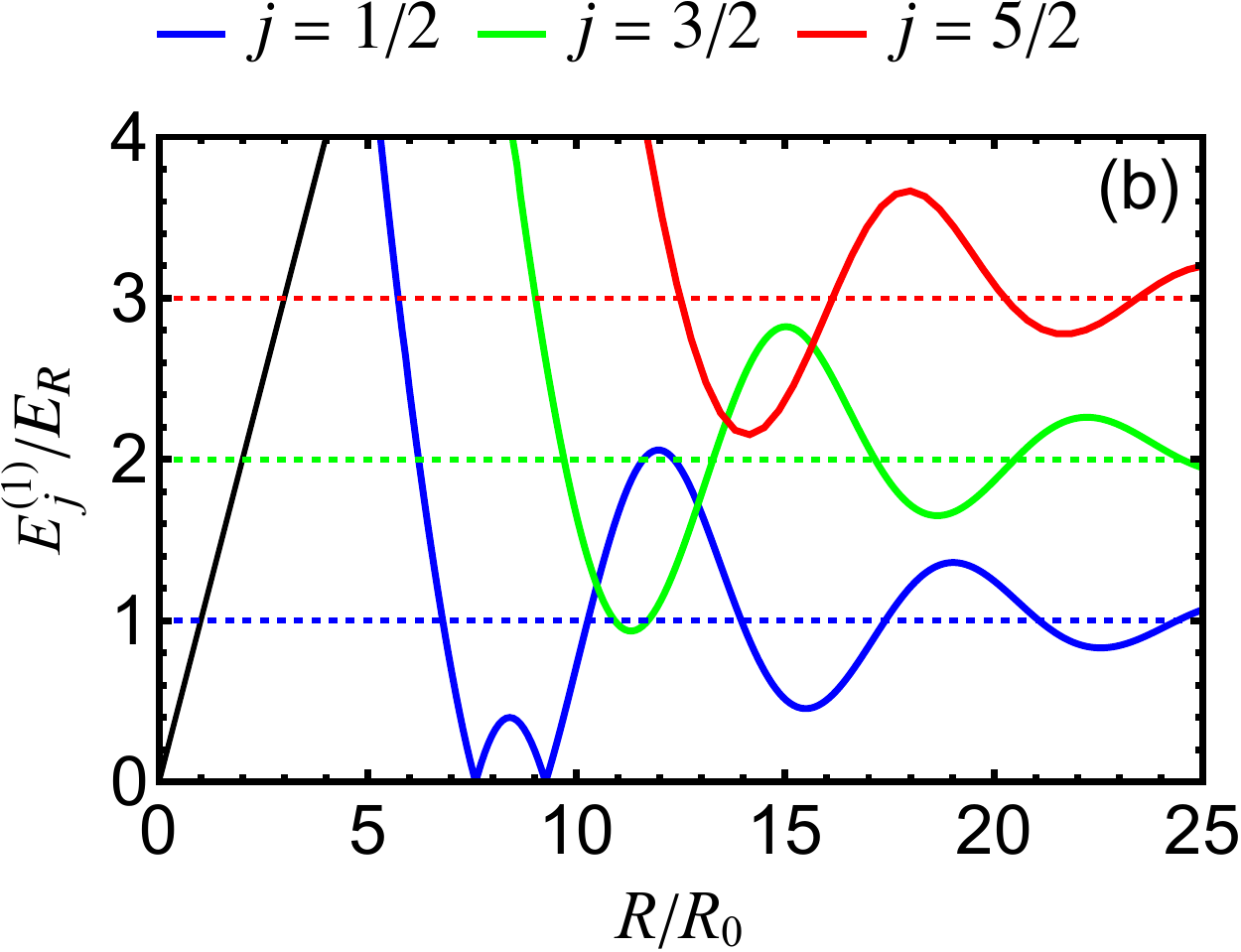}\hfill
\includegraphics[width=0.323\textwidth]{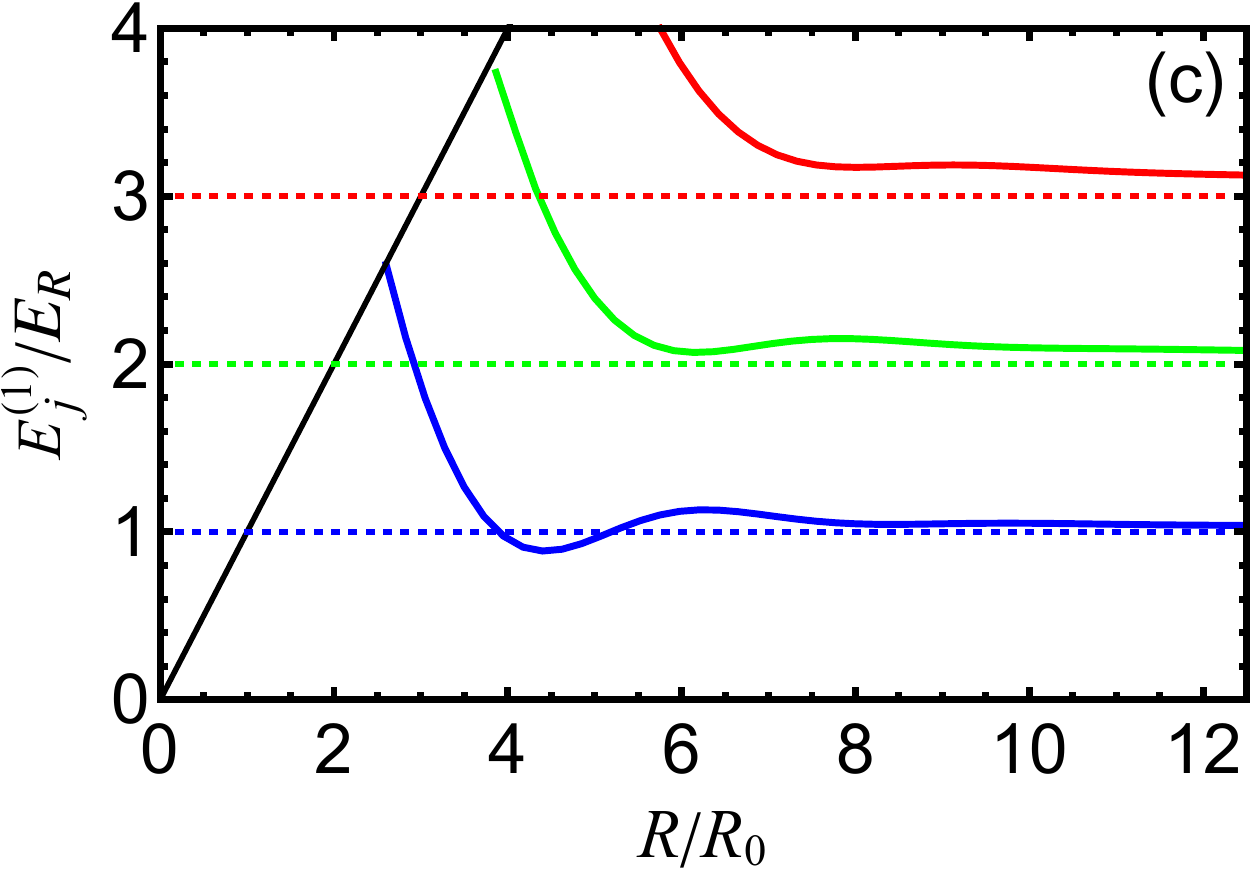}
\caption{\label{fig:topState}%
Energies $E^{(1)}_j$ of quantized electron levels created from
topologically protected surface states by a spherical hard-wall
mass-confinement potential with radius $R$. The solid blue (green,
red) curve in each panel corresponds to the state with
total-angular-momentum quantum number $j=1/2$ ($j=3/2$, $j=5/2$).
Here $E_R=\gamma/R$ is the size-quantisation energy scale and $R_0
= 2\gamma/|\Delta_0|$ the bulk Compton length. Results shown in panel
(a), (b) and (c) were obtained for fixed $\Delta_0/(\gamma k_\Delta)
= -1.2$, $-9.5$ and $-2.3$, respectively, which are values
representative of Bi$_2$Se$_3$, Bi$_2$Te$_3$ and Sb$_2$Te$_3$ (see
Table~\ref{tab:parameters} for these materials' band-structure
parameters). The thin black line indicates the bulk-material
conduction-band edge.}
\end{figure*}

Previous theoretical studies of TI nanoparticles have used atomistic
electronic-structure calculations~\cite{mal10}, approximate
descriptions for the surface states~\cite{imu12,tak13}, discretized
versions of effective $\kk\cdot\vek{p}$ models~\cite{imu12,sir16,
sir17,zir17}, and numerical exact diagonalization~\cite{neu15,dur16}
to discuss excitonic properties~\cite{mal10}, effects of
interactions~\cite{neu15} and disorder~\cite{dur16,sir17}, as well as
the emergence of unusual electromagnetic properties~\cite{sir16,
zir17}. Related works have considered idealized Dirac models in curved
spaces~\cite{abr02,lee09,par11} or subject to spherically symmetric
quantum confinement~\cite{lay11,pau13}. Here we obtain the exact
spectrum of quantised energies and associated quantum states for
nanoparticle confinement of charge carriers described by a spherically
symmetric version of the low-energy continuum-model Hamiltonian for
bulk TIs~\cite{zha09,liu10,bre18},
\begin{align}\label{eq:3DbulkHam}
H =&\left( \frac{\Delta_0}{2} + \frac{\gamma}{k_\Delta}\, \kk^2\right)
\tau_z \otimes \sigma_0 \nonumber \\[0.1cm] & + \gamma \left( k_z\,
\tau_x \otimes \sigma_z + k_-\, \tau_x \otimes \sigma_+ + k_+ \,
\tau_x \otimes \sigma_- \right) \, ,
\end{align}
comprising both pseudo-spin ($\tau$) and real-spin ($\sigma$) degrees
of freedom~\cite{notePauli}. In Eq.~(\ref{eq:3DbulkHam}), $\kk \equiv
(k_x, k_y, k_z) = -i \vek{\nabla}$ is the operator for the
single-electron wave vector, and $k_\pm = k_x \pm i\, k_y$. Further
details of the formalism are discussed in Sec.~\ref{sec:states}, and
Table~\ref{tab:parameters} provides the parameter values applicable to
a number of currently available TI materials. We use the remainder
of this Section to present a selection of key results.

\begin{table}[t]
\caption{\label{tab:parameters}
Values adopted for parameters in the spherically symmetric effective
bulk continuum-model Hamiltonian (\ref{eq:3DbulkHam}) to represent
currently available topological-insulator materials. The Compton
length $R_0\equiv 2\gamma/|\Delta_0|$ for bulk charge carriers is
indicative for the magnitude of the nanoparticle radius $R$ below
which the band structure becomes topologically trivial.}
\renewcommand{\arraystretch}{1.1}
\begin{tabular*}{\columnwidth}{lfccccc}
\hline \hline \rule{0pt}{2.5ex}
& $\gamma$ (eV\AA) & $\Delta_0$ (eV) & $k_\Delta$ (\AA$^{-1}$) &
$R_0$ (nm) \\ \hline
Bi$_2$Te$_3$\footnotemark[1] & $2.4$ & $-0.592$ & $0.026$ & $0.81$ \\
Sb$_2$Te$_3$\footnotemark[1] & $2.4$ & $-0.364$ & $0.066$ &
$1.3$ \\
Bi$_2$Se$_3$\footnotemark[1]\footnotetext{Ref.~\cite{nec16}
(parameters averaged over spatial anisotropies)} & $2.2$ & $-0.338$ &
$0.13$ & $1.3$ \\
HgTe\footnote{Refs.~\cite{nov05,kli15}} & $8.4$ & $-0.303$ & $0.09$ &
$5.5$ \\
Pb$_{0.81}$Sn$_{0.19}$Se\footnote{Ref.~\cite{ass17}} & $3.2$ &
$-0.025$ & $0.20$ & $26$ \\
Pb$_{0.54}$Sn$_{0.46}$Te\footnote{Ref.~\cite{ass16}} & $4.8$ &
$-0.030$ & $\gtrsim 0.25$ & $32$ \\ \hline \hline
\end{tabular*}
\end{table}

The nanoparticle confinement leads to the emergence of quantized
energies $E_j^{(n)}$ for single-electron states labeled by the
half-integer total-angular-momentum quantum number $j$ and a radial
quantum number $n$. For large-enough nanoparticle radius $R$, the
lowest positive-energy levels $E_j^{(1)}$ are below the bulk
conduction-band edge $\Delta_0/2$. These remnants of the topologically
protected surface states of the bulk system are pushed to higher
energies as $R$ is reduced and, one-by-one starting from the largest
$j$, disappear from the subgap energy range. This is illustrated for
the three lowest positive-energy levels in Fig.~\ref{fig:topState},
where we adopted the TI-material Compton length $R_0\equiv 2\gamma/
|\Delta_0|$ and the size-quantization energy scale $E_R\equiv \gamma
/R$ as natural units. The values for $\Delta_0/(\gamma k_\Delta)$ used
in the calculation of results shown in panels (a), (b) and (c) are
derived from band-structure parameters given in
Table~\ref{tab:parameters} for Bi$_2$Se$_3$, Bi$_2$Te$_3$ and
Sb$_2$Te$_3$, respectively. Clear deviations from the previously
determined~\cite{imu12,pau13} asymptotic large-$R$ behavior
$E_{|\lambda|}^{(1)}\sim \big( j + \frac{1}{2} \big) E_R$ occur over
a wider range of $R/R_0$ for larger $j$ and increasing $|\Delta_0|/
(\gamma k_\Delta)$. Bi$_2$Te$_3$ represents a case with strong
deviations in the form of large oscillations and a steep rise in
bound-state energy at the smallest values of $R/R_0$. The
Bi$_2$Te$_3$ nanoparticle also exhibits gaplessness at two values of
$R/R_0$, and there is a finite range of radii where the $j=3/2$ level
is the lowest-energy state and thus defines the TI-nanoparticle band
gap. Similar subband inversions occur between higher-$j$ levels. The
lowest energy level is pushed above the bulk-conduction-band edge for
$R < R_\mathrm{c} = 2.0\, R_0$ ($2.6\, R_0$, $5.0\, R_0$) in the
system representing a Bi$_2$Se$_3$ (Sb$_2$Te$_3$, Bi$_2$Te$_3$)
nanoparticle, marking the transition to a nontopological band
structure. Thus although $R_0$ sets the overall scale for the critical
radius below which the TI-nanoparticle spectrum becomes ordinary, the
actual value of $R_\mathrm{c}$ is typically several times larger.

The wave functions in position ($\rr$) space for
TI-nanoparticle-confined electrons can be written in the universal
form
\begin{widetext}
\begin{equation}\label{eq:univWF}
\Psi_{j m +}^{(n)}(\rr) = \frac{C_{j+}^{(n)}}{2} \begin{pmatrix}
\sqrt{\frac{j+m}{j}}\,\, Y_{j-\frac{1}{2}}^{m-\frac{1}{2}}(\theta,
\varphi) \,\, \phi_{j + \uparrow}^{(n)}(r) \\[0.3cm]
\sqrt{\frac{j+1-m}{j+1}}\,\, Y_{j+\frac{1}{2}}^{m-\frac{1}{2}}(\theta,
\varphi)\,\, \phi_{j - \uparrow}^{(n)}(r)\\[0.3cm]
\sqrt{\frac{j-m}{j}}\,\, Y_{j-\frac{1}{2}}^{m+\frac{1}{2}}(\theta,
\varphi)\,\, \phi_{j + \uparrow}^{(n)}(r) \\[0.3cm]
-\sqrt{\frac{j+1+m}{j+1}}\,\, Y_{j+\frac{1}{2}}^{m+\frac{1}{2}}
(\theta,\varphi)\,\, \phi_{j - \uparrow}^{(n)}(r) \end{pmatrix} \,\,
, \,\, \Psi_{j m -}^{(n)}(\rr) =\frac{C_{j-}^{(n)}}{2} \begin{pmatrix}
\sqrt{\frac{j+1-m}{j+1}}\,\, Y_{j+\frac{1}{2}}^{m-\frac{1}{2}}(\theta,
\varphi)\,\, \phi_{j + \uparrow}^{(n)}(r) \\[0.3cm]
\sqrt{\frac{j+m}{j}}\,\, Y_{j-\frac{1}{2}}^{m-\frac{1}{2}}(\theta,
\varphi)\,\, \phi_{j - \uparrow}^{(n)}(r) \\[0.3cm]
-\sqrt{\frac{j+1+m}{j+1}}\,\, Y_{j+\frac{1}{2}}^{m+\frac{1}{2}}
(\theta,\varphi)\,\, \phi_{j + \uparrow}^{(n)}(r) \\[0.3cm]
\sqrt{\frac{j-m}{j}}\,\, Y_{j-\frac{1}{2}}^{m+\frac{1}{2}}(\theta,
\varphi)\,\, \phi_{j - \uparrow}^{(n)}(r)
\end{pmatrix}
\end{equation}
\end{widetext}
associated with the ordered set of basis states~\cite{zha09,liu10,
bre18}
\begin{align}\label{eq:basis}
&\ket{P1_-^+,1/2} \equiv \ket{+\uparrow}\,\, , \,\, -i\ket{P2_+^-,1/2}
\equiv \ket{-\uparrow} \,\, , \nonumber \\  & \hspace{0.5cm}
\ket{P1_-^+,-1/2} \equiv \ket{+\downarrow}\,\, , \,\,
i\ket{P2_z^-,-1/2} \equiv \ket{-\downarrow}\,\, .
\end{align}
Here $Y_{j\pm\frac{1}{2}}^{m\pm\frac{1}{2}}(\theta,\varphi)$ are the
familiar spherical harmonics as defined in Ref.~\cite{sak11}. The
physically unimportant phase factors $C_{j\kappa}^{(n)}$ are given
below [see Eq.~(\ref{eq:prefact})]; they were chosen to make contact
with mathematical formalisms used to describe massless Dirac fermions
confined to move on the surface of the unit sphere~\cite{abr02}. The
functions $\phi_{j\tau \sigma}^{(n)}(r)$ are determined by solving a
radial Schr\"odinger equation as explained in greater detail in
Sec.~\ref{sec:states}. Noting that the spherical harmonics
$Y_{|\lambda|}^{\mu}$ acquire a factor $(-1)^{|\lambda|}$ under the
parity transformation~\cite{sak11}, and that the basis states
$\ket{+\, \sigma}$ ($\ket{-\, \sigma}$) are even (odd) under
parity~\cite{zha09,liu10,bre18}, the wave functions $\Psi_{j m
\pm}^{(n)}(\rr)$ are found to be parity eigenstates with eigenvalue
$(-1)^{j\mp\frac{1}{2}}$. The angular part of the wave functions from
Eq.~(\ref{eq:univWF}) has a structure analogous to that exhibited by
localized states in narrow-gap semiconductors~\cite{she67,efr98} and
spherically symmetric solutions of the Dirac equation~\cite{tha92}.

The explicit form of the wave functions (\ref{eq:univWF}) enables
straightforward calculation of the amplitude for optical transitions
between quantized electron states in a TI nanoparticle. In
Sec.~\ref{sec:optProp}, we derive the general expression
\begin{equation}\label{eq:envDip}
\vek{d} = e \, \rr\, \tau_0\otimes \sigma_0 + \frac{e\, R_0}{2}\,
\tau_y\otimes\vek{\sigma}
\end{equation}
for the optical-dipole operator acting in 4-dimensional
envelope-function space of combined pseudo-spin ($\tau$) and real-spin
($\sigma$) degrees of freedom~\cite{notePauli} describing electrons in
the TI material~\footnote{Within a conventional
nomenclature~\cite{hau04}, the term proportional to the position
vector $\rr$ (the TI-material Compton length $R_0$) in
Eq.~(\ref{eq:envDip}) is associated with intra-band (inter-band)
transitions. This jargon is somewhat misleading in the case of
narrow-gap (e.g., TI) materials~\cite{pau13}. More precisely, the
first (second) term on the r.h.s.\ of Eq.~(\ref{eq:envDip}) pertains
to dipole transitions mediated by the envelope part (the
Bloch-function basis) of the confined-electron states.}. The optical
matrix elements
\begin{equation}\label{eq:dipMatGen}
\vek{d}_{n\, j\, m\, \kappa}^{n' j' m' \kappa'} \equiv \int d^3 r \,\,
\big[ \Psi_{j' m' \kappa'}^{(n')}(\rr)\big]^\dagger  \, \vek{d} \,\,
\Psi_{j m \kappa}^{(n)}(\rr)\,\, ,
\end{equation}
for the transition from an initial state $\Psi_{j m \kappa}^{(n)}$ to
a final state $\Psi_{j' m' \kappa'}^{(n')}$ are explicitly found as
\begin{widetext}
\begin{subequations}\label{eq:transAmp}
\begin{eqnarray}
(d_x \pm i d_y)_{n\, j\, m\, \kappa}^{n' j' m' \kappa'} &=& e\,
\delta_{m', m\pm 1} \Bigg\{ \delta_{\kappa', \kappa} \sum_{\xi=\pm 1}
\frac{\sqrt{\left[ j+\frac{1}{2}+\xi (\frac{1}{2} \pm m)\right]\left[
j+\frac{1}{2}+\xi (\frac{3}{2}\pm m)\right]}}{2j +1+\xi} \left( \mp\xi
\, \mathcal{R}_{n\, j}^{n' j'}\pm i \kappa \, \frac{R_0}{2} \,
\mathcal{S}_{n\, j, -\kappa\xi}^{n' j', \kappa\xi}\right) \delta_{j',
j+\xi} \nonumber \\[0.2cm] && \hspace{-1cm} +\,\, \delta_{\kappa',
-\kappa}\,\, \frac{\sqrt{(j\mp m)(j+1 \pm m)}}{2j(j+1)}\left(
\mathcal{R}_{n\, j}^{n' j'} + i\kappa\, \frac{R_0}{2} \left[ j \,
\mathcal{S}_{n\, j, -\kappa}^{n' j', \kappa} + (j+1)\, \mathcal{S}_{n
\, j, \kappa}^{n' j', -\kappa} \right]\right) \delta_{j', j} \Bigg\}
\Big[ C_{j'\kappa'}^{(n')} \Big]^\ast C_{j\kappa}^{(n)} \,\, , \quad
\\[0.2cm]
(d_z)^{n' j' m' \kappa'}_{n\, j\, m\, \kappa} &=& e\, \delta_{m', m}\,
\Bigg\{ \delta_{\kappa', \kappa} \sum_{\xi=\pm 1}\, \frac{\sqrt{\left[
j + (1+\xi)\frac{1}{2} + m \right] \left[ j + (1+\xi) \frac{1}{2} -m
\right]}}{2 j+1+ \xi}\left( \mathcal{R}_{n\, j}^{n' j'} - i \kappa \xi
\, \frac{R_0}{2}\,\mathcal{S}_{n\, j, -\kappa\xi}^{n' j', \kappa\xi}
\right) \delta_{j', j+\xi}\nonumber \\[0.2cm] && \hspace{1cm} +\,\,
\delta_{\kappa', -\kappa} \,\, \frac{m}{2j(j+1)}\left[\mathcal{R}_{n\,
j}^{n' j'} + i\kappa\, \frac{R_0}{2} \left( j \, \mathcal{S}_{n \, j,
-\kappa}^{n' j', \kappa} + (j+1)\, \mathcal{S}_{n\, j, \kappa}^{n' j',
-\kappa} \right) \right] \delta_{j', j} \Bigg\} \Big[
C_{j'\kappa'}^{(n')} \Big]^\ast C_{j\kappa}^{(n)} \,\, .
\end{eqnarray}
\end{subequations}
For convenience, we have introduced the abbreviation
\begin{equation}\label{eq:effectRad}
\mathcal{R}_{n\, j}^{n' j'} = \frac{1}{2}\, \int_0^\infty dr \,\,\,
r^3 \, \left\{ \left[ \phi_{j' + \uparrow}^{(n')}(r) \right]^\ast
\phi_{j + \uparrow}^{(n)}(r) + \left[ \phi_{j' - \uparrow}^{(n')}(r)
\right]^\ast \phi_{j - \uparrow}^{(n)}(r) \right\}
\end{equation}
\end{widetext}
for the matrix element of the radial coordinate between initial and
final radial spinor wave functions from Eq.~(\ref{eq:univWF}), and the
symbols
\begin{equation}\label{eq:radOlap}
\mathcal{S}_{n\, j, \tau}^{n' j', \tau'} = \int_0^\infty dr \,\,\, r^2
\,\, \left[ \phi_{j' \tau' \uparrow}^{(n')}(r) \right]^\ast \phi_{j
\tau \uparrow}^{(n)}(r)
\end{equation}
are overlap integrals of radial-wave-function components.

The dependence of matrix elements given in Eqs.~(\ref{eq:transAmp})
on the angular-momentum quantum numbers of the initial and final
states embodies selection rules for optical transitions between energy
levels in TI nanoparticles. For states with same-type parity ($\kappa'
=\kappa$), the rules $j'-j = \pm 1$ and $m'-m=0, \pm 1$ are identical
to those governing atomic transitions~\cite{gre98}. However, there is
also a previously overlooked~\cite{pau13} possibility for transitions
between states that are of opposite type with respect to parity (i.e.,
$\kappa'=-\kappa$), and these follow the unconventional rule $j'-j =
0$ while still adhering to $m'-m=0, \pm 1$. As it turns out, this
unconventional type of transition actually determines the low-energy
threshold for optical absorption and emission in undoped TI
nanoparticles.

\begin{figure}[t]
\includegraphics[width=0.95\columnwidth]{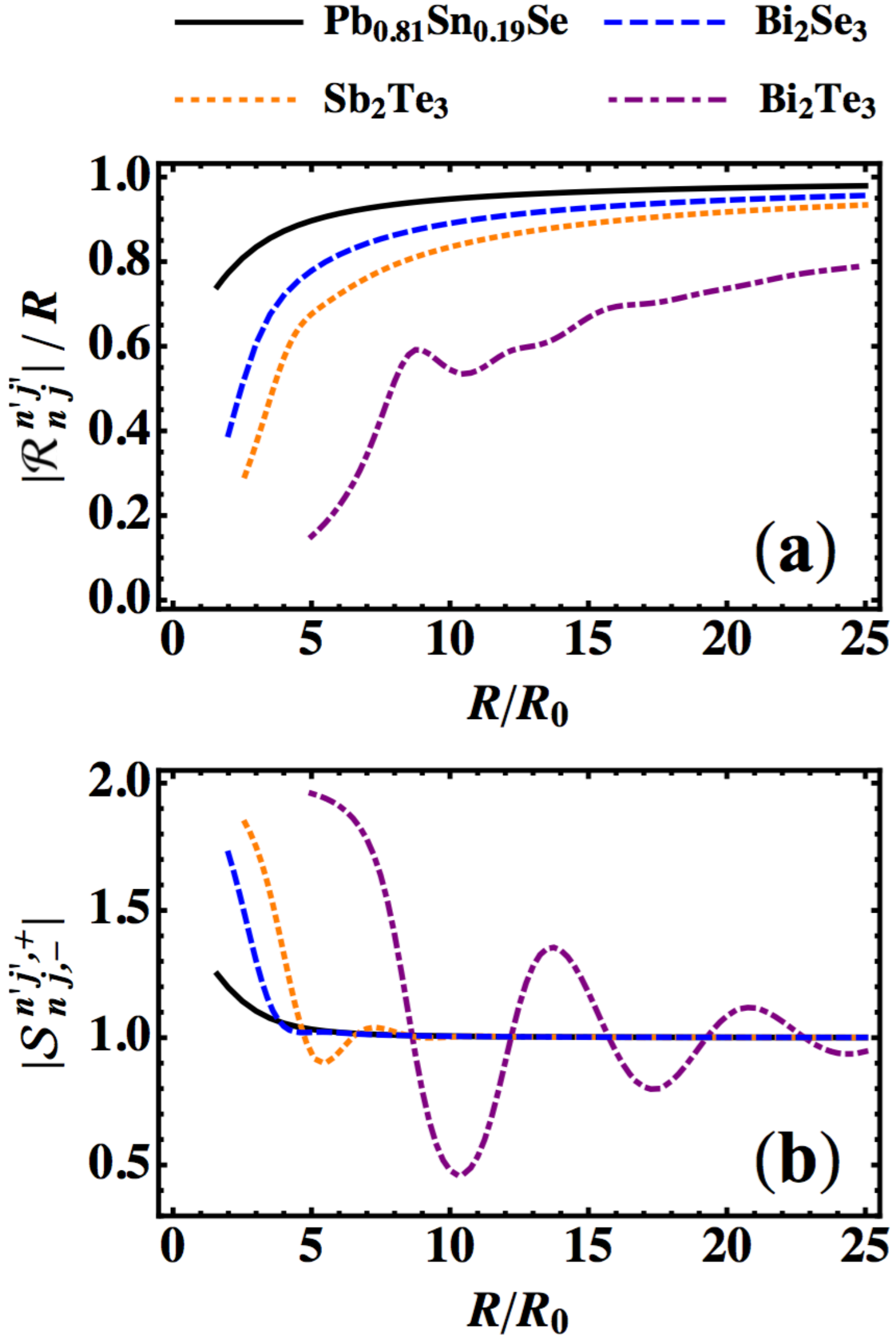}
\caption{\label{fig:radMat}%
Dependence of optical-transition amplitudes on TI-nanoparticle size.
Panel (a) [(b)] shows the magnitude of the radial-coordinate matrix
element $\mathcal{R}_{n\, j}^{n' j'}$ $\big[$the overlap integral
$\mathcal{S}_{n\, j, -}^{n' j',+}\big]$ for $j=j'=1/2$ and $n=-1$,
$n'=1$, which typically corresponds to the transition across the
intrinsic nanoparticle band gap, as a function of the nanoparticle
radius $R$ over the range where the lowest confined electron level is
still below the bulk-band-gap edge. Here $R_0 = 2\gamma/|\Delta_0|$ is
the bulk Compton length, and the black solid (blue dashed, orange
dotted, purple dot-dashed) curve corresponds to $\Delta_0/(\gamma
k_\Delta) = -0.039$ ($-1.2$, $-2.3$, $-9.5$,), which is the value
representative of Pb$_{0.81}$Sn$_{0.19}$Se (Bi$_2$Se$_3$,
Sb$_2$Te$_3$, Bi$_2$Te$_3$). From results presented in panel~(b),
$\big|\mathcal{S}_{n\, j, +}^{n' j',-}\big| \equiv 2 - \big|
\mathcal{S}_{n\, j, -}^{n' j',+}\big|$ can be inferred.}
\end{figure}

The effect of the radial confinement on dipole-allowed single-electron
transitions in TI nanoparticles is contained in the matrix elements
$\mathcal{R}_{n\, j}^{n' j'}$ and the overlap integrals
$\mathcal{S}_{n\, j, \tau}^{n' j', \tau'}$ defined in
Eqs.~(\ref{eq:effectRad}) and (\ref{eq:radOlap}), respectively. In the
large-size limit $R\gg R_0$, we find $|\mathcal{R}_{n\, j}^{n' j'}|
\sim R$ and $|\mathcal{S}_{n\, j, \mp}^{n' j',\pm}|\sim 1$, reflecting
the surface-localized massless-Dirac-fermion character of low-energy
states in this regime. However, as illustrated in
Figs.~\ref{fig:radMat}(a) and \ref{fig:radMat}(b), significant
deviations from this expected limiting behavior occur in the range of
small-to-medium nanoparticle sizes $R\gtrsim R_0$. These deviations
become more sizable, and also occur over a greater size range, for
larger values of the TI-material parameter $-\Delta_0/(\gamma
k_\Delta)$.

The remainder of this Article is structured as follows. In
Section~\ref{sec:states}, the continuum-model description of 3D
topological insulators based on Hamiltonian (\ref{eq:3DbulkHam}) is
used to obtain the spectrum of quantized energies and associated
quantum states for a spherical hard-wall mass-confinement potential.
Here we derive the general form of the confined-electron eigenspinors
given in Eq.~(\ref{eq:univWF}) and also find the radial wave functions
$\phi_{j \tau \sigma}^{(n)}(r)$. The formalism for describing optical
properties of TI nanoparticles is developed in
Section~\ref{sec:optProp}. Transitions can be mediated by both the
envelope-function part of electron wave functions and also by the
Bloch-function basis states. We treat these contributions on the same
footing and find a compact expression for a total optical-dipole
operator that subsumes both of them [Eq.~(\ref{eq:envDip})]. Applying
the result from Secs.~\ref{sec:states} and \ref{sec:optProp}, we
obtain the optical-transition matrix elements given in
Eqs.~(\ref{eq:transAmp}). These form the basis for discussing
quantum-size effects in TI nanoparticles, focusing especially on their
ramifications for optical spectroscopy. Our conclusions are presented
in Section~\ref{sec:outlook}.

\section{Electron states in spherically symmetric TI nanostructures}
\label{sec:states}

Our starting point is the effective continuum-model Hamiltonian
(\ref{eq:3DbulkHam}), which is associated with the set of basis
states given in Eq.~(\ref{eq:basis}). Throughout this work, we label
basis states $\ket{\tau\,\sigma}$ in terms of pseudo-spin $\tau\in
\{+,-\}$ and real spin $\sigma\in\{\uparrow,\downarrow
\}$~\footnote{The degree of freedom associated with the quantum
number $\sigma$ is the projection of a real spin-$1/2$ angular
momentum, but this is not the actual electron spin~\cite{bre18}.}. The
parameter $\gamma$
characterizes the inter-band coupling, and $\Delta_0$ is the energy
gap of the bulk-TI material. The term involving $k_\Delta$ quantifies
remote-band contributions that also provide a natural regularization
of the leading Dirac-fermion-like dispersion that is essential for
keeping the theory well-defined in the ultra-relativistic limit
$\Delta_0\to 0$. To facilitate an instructive analytical treatment,
the often quite sizeable deviations from spherical symmetry and
particle-hole symmetry in the TI-material bulk dispersions are
omitted from the Hamiltonian (\ref{eq:3DbulkHam}), and materials
parameters given in Table~\ref{tab:parameters} correspond to spatial
averages.

We first consider the general spherical-confinement problem in
Subsection~\ref{sec:genConf}. Using a wave-function \textit{Ansatz\/}
involving the known eigenstates of a massless Dirac fermion moving on
the surface of the unit sphere~\cite{abr02}, we obtain a
one-dimensional radial Schr\"odinger equation. The latter's solution
for a hard-wall mass-confinement potential is presented in
Subsection~\ref{sec:massConf}.

\subsection{Separation Ansatz and radial Schr\"odinger equation}
\label{sec:genConf}

We envision the TI nanoparticle to be defined by a spherically
symmetric mass-confinement potential
\begin{equation}\label{eq:TINPSE}
H_V = V(r) \, \tau_z \otimes \sigma_0 \quad .
\end{equation}
Then the electronic-bound-state Schr\"odinger equation $\left[ H + H_V
\right] \Psi = E\, \Psi$ is most conveniently solved by decoupling
angular and radial motions. Expressing $H$ in terms of the usual
spherical coordinates $\rr\equiv (r, \theta, \varphi)$ and using the
transformation~\cite{imu12}
\begin{equation}
\mathcal{U}(\theta,\varphi) = \tau_0\otimes \exp \left( -i\,
\frac{\varphi}{2} \, \sigma_z \right) \exp \left( -i\,
\frac{\theta}{2}\, \sigma_y \right) \quad ,
\end{equation}
it is straightforward to obtain
\begin{equation}
H = \mathcal{U}(\theta,\varphi) \left( \mathcal{H}^\perp +
\mathcal{H}^\parallel \right) \mathcal{U}^\dagger(\theta,\varphi)
\quad ,
\end{equation}
with Hamiltonians associated with radial and angular motion,
respectively, given by
\begin{widetext}
\begin{subequations}
\begin{eqnarray}
\mathcal{H}^\perp &=& \left[ \frac{\Delta_0}{2} -
\frac{\gamma}{k_\Delta} \, \frac{1}{r^2} \, \partial_r \left( r^2\,
\partial_r \right) \right] \tau_z \otimes \sigma_0 - i\, \gamma \left(
\partial_r + \frac{1}{r} \right) \tau_x \otimes \sigma_z \quad ,
\\[0.2cm]
\mathcal{H}^\parallel &=& \frac{\gamma}{r} \left[ D\, \tau_x \otimes
\sigma_+ + D^\dagger \, \tau_x\otimes \sigma_- + \frac{1}{k_\Delta r}
\left( D D^\dagger\, \tau_z\otimes \pi_+ + D^\dagger D\, \tau_z\otimes
\pi_- + i\, D\, \tau_z \otimes \sigma_+ - i\, D^\dagger\, \tau_z
\otimes \sigma_- \right) \right] \, .
\end{eqnarray}
\end{subequations}
\end{widetext}
Here $\pi_\pm = (\sigma_0 \pm i\sigma_z)/2$ are projection operators
on the eigenstates of $\sigma_z$ in real-spin space, and the operators
\begin{subequations}
\begin{eqnarray}
D &=& -i \left( \partial_\theta + \frac{\cot\theta}{2} - \frac{i\,
\partial_\varphi}{\sin\theta} \right)  \quad , \\[0.1cm]
D^\dagger &=& - i\left( \partial_\theta + \frac{\cot\theta}{2} +
\frac{i\, \partial_\varphi}{\sin\theta} \right)
\end{eqnarray}
\end{subequations}
appear in the Dirac operator on the unit sphere~\cite{abr02}.

The spherical mass-confinement potential $H_V$ trivially commutes with
the transformation $\mathcal{U}(\theta,\varphi)$, i.e., $\mathcal{H}_V
\equiv \mathcal{U}^\dagger(\theta,\varphi)\, H_V \,\, \mathcal{U}
(\theta,\varphi) = H_V$. This feature, together with the structure of
the angular-motion Hamiltonian $\mathcal{H}^\parallel$, motivates the
separation \textit{Ansatz}
\begin{equation}\label{eq:Spinor}
\Psi(r, \theta, \varphi) = \mathcal{U}(\theta,\varphi)\,\,
\frac{\ee^{i m \varphi}}{\sqrt{4\pi}} \begin{pmatrix}
\alpha_{\lambda m}(\theta)\, \phi_{|\lambda| + \uparrow}(r) \\[2pt]
\alpha_{\lambda m}(\theta)\, \phi_{|\lambda| - \uparrow}(r) \\[2pt]
\beta_{\lambda m}(\theta)\, \phi_{|\lambda| + \downarrow}(r) \\[2pt]
\beta_{\lambda m}(\theta)\, \phi_{|\lambda| - \downarrow}(r)
\end{pmatrix} \,\, ,
\end{equation}
where the functions $\alpha_{\lambda m}(\theta)$ and $\beta_{\lambda
m}(\theta)$ are the polar-angle-dependent entries in eigenspinors of a
massless Dirac fermion moving on a spherical shell~\cite{abr02,imu12}.
Using the fact that they satisfy the set of relations
\begin{subequations}
\begin{eqnarray}
&D\, \beta_{\lambda m}(\theta) = \lambda\, \alpha_{\lambda m}(\theta)
\,\, , \,\,
D^\dagger\, \alpha_{\lambda m}(\theta) = \lambda\, \beta_{\lambda m}
(\theta) \,\, , \quad \\[0.1cm]
&\alpha_{\lambda m} = \left( \frac{\lambda}{|\lambda|}
\right)^{\frac{1}{2}} \alpha_{|\lambda| m} \,\, , \,\,
\beta_{\lambda m} = \left( \frac{\lambda}{|\lambda|}
\right)^{-\frac{1}{2}} \beta_{|\lambda| m} \,\, , \quad
\end{eqnarray}
\end{subequations}
we obtain the radial Schr\"odinger equation
\begin{equation}\label{eq:radialSE}
\left[ \mathcal{H}^\perp_{|\lambda|} + \mathcal{H}_V \right] 
\Phi_{|\lambda|}(r) = E_{|\lambda|} \, \Phi_{|\lambda|}(r)
\end{equation}
for the spinor $\Phi_{|\lambda|} = \big( \phi_{|\lambda| + \uparrow},
\phi_{|\lambda| - \uparrow}, \phi_{|\lambda| + \downarrow},
\phi_{|\lambda| - \downarrow} \big)^T$, with the free radial
Hamiltonian given by
\begin{widetext}
\begin{equation}\label{eq:radialHam}
\mathcal{H}^\perp_{|\lambda|} =\left\{ \frac{\Delta_0}{2} +
\frac{\gamma}{k_\Delta} \frac{1}{r^2} \left[ |\lambda|^2 - \partial_r
\left( r^2\, \partial_r \right) \right] \right\} \tau_z \otimes
\sigma_0 - i\, \gamma \, \left( \partial_r + \frac{1}{r} \right)
\tau_x \otimes \sigma_z + \frac{\gamma}{r} \, |\lambda| \left( \tau_x
\otimes \sigma_x - \frac{1}{k_\Delta r}\, \tau_z \otimes \sigma_y
\right) \,\, .
\end{equation}
\end{widetext}

As previously discussed, e.g., in Ref.~\cite{abr02}, $|\lambda|$
is related to the half-integer eigenvalue $j$ of total angular
momentum $\vek{J} \equiv \left(\vek{L} + \frac{\hbar}{2}\,
\vek{\sigma} \right)$ via $|\lambda| = j+\frac{1}{2}$ and therefore
must be a nonzero integer. The quantum numbers $m=-j, -j+1, \dots j$
are the eigenvalues of the angular-momentum projection and, thus, also
assume half-integer values. With this additional insight, solution of
the radial Schr\"odinger equation (\ref{eq:radialSE}) can proceed for
any specific potential $V(r)$. Because of the one-to-one
correspondence between values of $|\lambda|$ and $j$, we can use
both interchangeably as labels for radial spinor wave functions;
$\Phi_{|\lambda|}(r)\equiv \Phi_j(r)$, and its components
$\phi_{|\lambda| \tau \sigma}(r)\equiv \phi_{j \tau \sigma}(r)$. In
the following Subsection~\ref{sec:massConf}, we consider the special
case of a hard-wall confinement. Using results obtained there as
input, the electronic eigenstates are found to be of the form
\begin{widetext}
\begin{subequations}
\begin{eqnarray}\label{eq:aState}
\Psi_{\lambda m}^{(n_\mathrm{a})}(\rr) &=& \mathcal{U}(\theta,
\varphi) \,\, \frac{\ee^{i m \varphi}}{\sqrt{4\pi}}\, \left\{
\begin{pmatrix} \alpha_{\lambda m}(\theta) \\ 0 \\ i\,\beta_{\lambda
m}(\theta) \\ 0 \end{pmatrix} \phi_{j + \uparrow}^{(n_\mathrm{a})} (r)
+ \begin{pmatrix} 0 \\ \alpha_{\lambda m}(\theta) \\ 0 \\ -i\,
\beta_{\lambda m}(\theta) \end{pmatrix} \phi_{j -
\uparrow}^{(n_\mathrm{a})}(r) \right\}\quad ,\\
\Psi_{\lambda m}^{(n_\mathrm{b})}(\rr) &=& \mathcal{U}(\theta,
\varphi) \,\, \frac{\ee^{i m \varphi}}{\sqrt{4\pi}}\, \left\{
\begin{pmatrix} \alpha_{\lambda m}(\theta) \\ 0 \\ -i\, \beta_{\lambda
m}(\theta) \\ 0 \end{pmatrix} \phi_{j + \uparrow}^{(n_\mathrm{b})} (r)
+ \begin{pmatrix} 0 \\ \alpha_{\lambda m}(\theta) \\ 0 \\ i\,
\beta_{\lambda m}(\theta) \end{pmatrix} \phi_{j -
\uparrow}^{(n_\mathrm{b})} (r) \right\} \quad .
\label{eq:bState}
\end{eqnarray}
\end{subequations}
\end{widetext}
Here $n_\mathrm{a}$ and $n_\mathrm{b}$ are the radial-quantum-number
labels for two sets of energy levels arising from two distinct secular
equations. Adapting the approach presented in Ref.~\cite{abr02},
it is possible to express the wave functions (\ref{eq:aState}) and
(\ref{eq:bState}) in terms of spherical harmonics. This yields the
results presented in Eq.~(\ref{eq:univWF}) above, where
\begin{subequations}
\begin{align}
\Psi_{\lambda m}^{(n)}(\rr) \equiv \Psi_{j m +}^{(n)}(\rr) \quad &
\mbox{for $n\in \{n_\mathrm{a}\}$ and $\lambda>0$} \nonumber
\\[-0.1cm] & \mbox{or $n\in \{n_\mathrm{b}\}$ and $\lambda<0$}\quad ,
\\ \Psi_{\lambda m}^{(n)}(\rr) \equiv \Psi_{j m -}^{(n)}(\rr) \quad &
\mbox{for $n\in \{n_\mathrm{a}\}$ and $\lambda<0$} \nonumber
\\[-0.1cm] & \mbox{or $n\in \{n_\mathrm{b}\}$ and $\lambda>0$} \quad ,
\end{align}
\end{subequations}
and prefactors given by
\begin{equation}\label{eq:prefact}
C_{j\kappa}^{(n)} = \left\{ \begin{array}{cl} i^j & \mbox{$n\in
\{n_\mathrm{a}\}$, $\kappa=+$ or $n\in\{n_\mathrm{b} \}$, $\kappa=-$}
\\[0.1cm] -i^{j+1} & \mbox{$n\in \{n_\mathrm{a}\}$, $\kappa=-$ or
$n\in\{n_\mathrm{b}\}$, $\kappa=+$} \end{array} \right. .
\end{equation}

Our route to obtain (\ref{eq:univWF}) via the \textit{Ansatz\/}
(\ref{eq:Spinor}) is inspired by a particular way of solving the
Dirac equation on a spherical shell~\cite{abr02,imu12,neu15}, but this
method is far from unique. An equivalent radial Schr\"odinger
equation and eigenstates having the same angular-coordinate
dependences as those in (\ref{eq:univWF}) should transpire, e.g., from
writing \textit{Ans\"atze} for eigenstates of the original Hamiltonian
(\ref{eq:3DbulkHam}) in terms of spinor spherical
harmonics~\cite{var88} and applying the mathematical identities
satisfied by them~\cite{szm07}. 

\subsection{Solution of the radial Schr\"odinger equation for a
hard-wall mass confinenment}
\label{sec:massConf}

Solution of the radial Schr\"odinger equation (\ref{eq:radialSE}) for
a hard-wall mass confinement defining a TI nanoparticle with radius
$R$ requires obtaining eigenstates of ${\mathcal H}^\perp_{j+
\frac{1}{2}}$ from Eq.~(\ref{eq:radialHam}) and forming linear
combinations of these for fixed energy $E$ that satisfy the hard-wall
boundary condition at $r=R$. The secular equation associated with the
linear system of equations for the superposition coefficients yields
the bound-state energies.

There is a set of propagating-wave-like eigenstates for ${\mathcal
H}^\perp_{j+\frac{1}{2}}$ that have the general form
\begin{subequations}\label{eq:perpD}
\begin{eqnarray}
\Phi^{(\mathrm{D})}_{j, \mathrm{a}}(r) &=& \begin{pmatrix}
j_{j-\frac{1}{2}}(k\, r) \\[2pt] i\, \gamma_k \, j_{j+\frac{1}{2}} (k
\, r) \\[2pt] i\, j_{j-\frac{1}{2}}(k\, r) \\[2pt] \gamma_k \, j_{j+
\frac{1}{2}}(k\, r) \end{pmatrix} \quad , \\[0.07cm]
\Phi^{(\mathrm{D})}_{j, \mathrm{b}}(r) &=& \begin{pmatrix}
j_{j+\frac{1}{2}}(k\, r) \\[2pt] -i\, \gamma_k \, j_{j-\frac{1}{2}}
(k\, r) \\[2pt] -i\, j_{j+\frac{1}{2}}(k\, r) \\[2pt] \gamma_k \,
j_{j-\frac{1}{2}}(k\, r)
\end{pmatrix} \quad ,
\end{eqnarray}
\end{subequations}
where $j_\nu(\cdot)$ are spherical Bessel functions of the first
kind~\cite{abr64} (not to be confused with the quantum number $j$ of
total angular momentum), $k>0$ is a real number, and normalization
factors have been omitted because they get absorbed into the
coefficients in the bound-state linear combination. The energy
eigenvalues are given by
\begin{equation}
{\tilde E}^{(\mathrm{D})}_\pm(k) = \pm \sqrt{{\tilde k}^4 + \left( 1
+ \tilde \Delta_0\right){\tilde k}^2 + \frac{{\tilde\Delta}_0^2}{4}}
\quad ,
\end{equation}
and the parameter entering corresponding eigenspinors can be found
from the general expression
\begin{equation}\label{eq:gK}
\gamma_k = \mathrm{sgn}(E) \, \sqrt{\frac{2 \tilde E - \tilde \Delta_0
- 2 {\tilde k}^2}{2 \tilde E + \tilde\Delta_0 + 2 {\tilde k}^2}}
\quad .
\end{equation}
Here we started using natural units for energies and wave numbers via
the definitions $\tilde k = k/k_\Delta$ and $\tilde E = E/(\gamma
k_\Delta)$. The particular form of the eigenstates given in
Eqs.~(\ref{eq:perpD}) is reminiscent of confined-Dirac-particle
states~\cite{alb96,lay11,gio18}. The absence of spherical Bessel
functions of the second kind from the wave-function \textit{Ansatz\/}
(\ref{eq:perpD}) is dictated by the necessity of wave functions to be
well-behaved at $r=0$.

There exists another set of eigenstates for ${\mathcal H}^\perp_{j+
\frac{1}{2}}$ that are evanescent in character;
\begin{subequations}\label{eq:perpB}
\begin{eqnarray}
\Phi^{(\mathrm{B})}_{j, \mathrm{a}}(r) &=& \frac{1}{\sqrt{q\, r}}
\begin{pmatrix} \bar\gamma_q\, I_j(q\, r) \\[0.1cm] i\, I_{j+1}(q\, r)
\\[0.1cm] i\, \bar \gamma_q\, I_j(q\, r) \\[0.1cm] I_{j+1}(q\, r)
\end{pmatrix} \quad , \\[0.2cm]
\Phi^{(\mathrm{B})}_{j, \mathrm{b}}(r) &=& \frac{1}{\sqrt{q\, r}}
\begin{pmatrix} -\bar\gamma_q\, I_{j+1}(q\, r) \\[0.1cm] -i\, I_j(q\,
r) \\[0.1cm] i \, \bar\gamma_q\, I_{j+1}(q\, r) \\[0.1cm] I_j(q\, r)
\end{pmatrix} \quad ,
\end{eqnarray}
\end{subequations}
where the spinor entries are modified spherical Bessel functions of
the first kind~\cite{abr64}, and $q>0$ is real. The associated energy
eigenvalues are
\begin{equation}
{\tilde E}^{(\mathrm{B})}_\pm(q) = \pm \sqrt{{\tilde q}^4 - \left( 1
+ \tilde\Delta_0\right){\tilde q}^2 + \frac{{\tilde\Delta}_0^2}{4}}
\quad ,
\end{equation}
and
\begin{equation}\label{eq:BgQ}
\bar\gamma_q = \sqrt{\frac{2 {\tilde q}^2  - \tilde\Delta_0 - 2
\tilde E}{2 {\tilde q}^2 - \tilde\Delta_0 + 2\tilde E}} \quad .
\end{equation}
The consideration of such evanescent free-particle eigenstates in the
context of BHZ-type model Hamiltonians~\cite{ber06,zho08} is required
for mathematical consistency, but their physical significance is
limited~\cite{whi81,sch85,kli18}.

Having determined the eigenstates of ${\mathcal H}^\perp_{j+
\frac{1}{2}}$, we can make an \textit{Ansatz\/} for the wave function
of hard-wall-confined electrons in the TI nanoparticle in the form of
superpositions
\begin{eqnarray}\label{eq:Ansatz}
\Phi^{(n)}_j(r) &=& c_{n,j,\mathrm{a}}^{(\mathrm{D})}\,
\Phi^{(\mathrm{D})}_{j,\mathrm{a}}(r) + c_{n,j,
\mathrm{a}}^{(\mathrm{B})}\, \Phi^{(\mathrm{B})}_{j,\mathrm{a}}(r)
\nonumber \\[0.1cm] && \hspace{0.5cm} +\, c_{n,j,
\mathrm{b}}^{(\mathrm{D})}\, \Phi^{(\mathrm{D})}_{j,\mathrm{b}}(r) +
c_{n,j,\mathrm{b}}^{(\mathrm{B})} \, \Phi^{(\mathrm{B})}_{j,
\mathrm{b}}(r) \quad
\end{eqnarray}
of such eigenstates for fixed energy $E$ and require that the boundary
condition $\Phi^{(n)}_j(R) = 0$ is satisfied. The associated system of
linear equations for the coefficients $c_{n,j,
\mathrm{a/b}}^{(\mathrm{D},\mathrm{B})}$ has nontrivial solutions only
if one of the secular equations
\begin{equation}\label{eq:secu}
\gamma_k\, \bar\gamma_q = \left\{ \begin{array}{cl}
\frac{I_{j+1}(q R)}{I_j(q R)} \, \frac{j_{j-\frac{1}{2}}(k R)}{j_{j+
\frac{1}{2}}(k R)} & \quad\mbox{case a}, \\[0.3cm] -\frac{I_j(q
R)}{I_{j+1}(q R)} \, \frac{j_{j+\frac{1}{2}}(k R)}{j_{j-\frac{1}{2}}(k
R)} & \quad\mbox{case b}, \end{array} \right.
\end{equation}
is fulfilled. In Eq.~(\ref{eq:secu}), the quantities $k$, $q$,
$\gamma_k$ and $\bar\gamma_q$ are functions of $E$, given explicitly
by~\footnote{We find the expressions for $\tilde k$ and $\tilde q$ by
solving for these in the relations ${\tilde E}^{(\mathrm{D})}_\pm(k)
= \tilde E$ and ${\tilde E}^{(\mathrm{B})}_\pm(q) = \tilde E$,
respectively. Then $\gamma_k$ and $\bar\gamma_q$ are obtained by
inserting the expressions found for $\tilde k$ and $\tilde q$ into
Eqs.~(\ref{eq:gK}) and (\ref{eq:BgQ}), respectively.}
\begin{subequations}
\begin{eqnarray}
\tilde k &=& \left[ \frac{1}{2} \left( \sqrt{(1+ {\tilde\Delta}_0)^2
+ 4 {\tilde E}^2 - {\tilde \Delta}_0^2} - 1 - {\tilde\Delta}_0 \right)
\right]^{\frac{1}{2}} , \\[0.2cm]
\tilde q &=& \left[ \frac{1}{2} \left( \sqrt{(1 + {\tilde\Delta}_0)^2
+ 4 {\tilde E}^2 - {\tilde\Delta}_0^2} + 1 + {\tilde\Delta}_0 \right)
\right]^{\frac{1}{2}} , \\[0.2cm]
\gamma_k &=& \mathrm{sgn}(\tilde E)\, \left[\frac{2 \tilde E -
\sqrt{(1 + {\tilde\Delta}_0)^2 + 4 \tilde E^2 - {\tilde\Delta}_0^2} +
1}{2\tilde E + \sqrt{(1 + {\tilde\Delta}_0)^2 + 4 \tilde E^2 - {\tilde
\Delta}_0^2} - 1}\right]^{\frac{1}{2}} , \quad \nonumber \\ \\[0.2cm]
{\bar\gamma}_q &=& \left[ \frac{\sqrt{(1 + {\tilde\Delta}_0)^2 + 4
\tilde E^2 - {\tilde\Delta}_0^2} + 1 - 2 \tilde E}{\sqrt{(1 + {\tilde
\Delta}_0)^2 + 4 \tilde E^2 - {\tilde\Delta}_0^2} + 1 + 2 \tilde E}
\right]^{\frac{1}{2}} .
\end{eqnarray}
\end{subequations}
Solution of Eq.~(\ref{eq:secu}) for cases a and b yields associated
series of quantized energy values $E^{(n_\mathrm{a})}_j$ and
$E^{(n_\mathrm{b})}_j$, respectively. By construction, these satisfy
$\big| E^{(n_\mathrm{a/b})}_j\big| \ge |\Delta_0|/2$. The coefficients
in the superposition (\ref{eq:Ansatz}) of corresponding bound states
are straightforwardly found to be
\begin{subequations}
\begin{eqnarray}
c_{n_\mathrm{a},j,\mathrm{a}}^{(\mathrm{D})} &=& N_{n_\mathrm{a},j}
\, , \, c_{n_\mathrm{a},j,\mathrm{a}}^{(\mathrm{B})} = N_{n_\mathrm{a}
,j} \, \xi_{n_\mathrm{a},j,\mathrm{a}} \, , \nonumber \\ &&
\hspace{2.5cm} c_{n_\mathrm{a},j,\mathrm{b}}^{(\mathrm{D})} =
c_{n_\mathrm{a},j,\mathrm{b}}^{(\mathrm{B})} = 0 \,\, , \\
c_{n_\mathrm{b},j,\mathrm{b}}^{(\mathrm{D})} &=& N_{n_\mathrm{b},j} \,
,\, c_{n_\mathrm{b},j,\mathrm{b}}^{(\mathrm{B})} = N_{n_\mathrm{b},j}
\, \xi_{n_\mathrm{b},j,\mathrm{b}} \, , \nonumber \\ &&
\hspace{2.5cm} c_{n_\mathrm{b},j,\mathrm{a}}^{(\mathrm{D})} =
c_{n_\mathrm{b},j,\mathrm{a}}^{(\mathrm{B})} = 0 \,\, , \quad
\end{eqnarray}
with normalisation factors $N_{n_\mathrm{a/b},j}$ and the parameters
\begin{eqnarray}
\xi_{n_\mathrm{a},j,\mathrm{a}} &=& - \left. \frac{\gamma_k\, \sqrt{q
R}\, j_{j+\frac{1}{2}}(k R)}{I_{j+1}(q R)}\right|_{E =
E^{(n_\mathrm{a})}_j} \quad , \\[0.2cm]
\xi_{n_\mathrm{b},j,\mathrm{b}} &=& - \left. \frac{\gamma_k\,\sqrt{q
R}\, j_{j-\frac{1}{2}}(k R)}{I_j(q R)} \right|_{E =
E^{(n_\mathrm{b})}_j} \quad .
\end{eqnarray}
\end{subequations}

Motivated by the well-known~\cite{vol85,ber06,zho08,gio18} existence
of evanescent states having $|E| < |\Delta_0|/2$ in the topological
regime~\footnote{A different type of intrinsic
semiconductor-nanoparticle sub-gap state discussed in
Ref.~\cite{ser99} has its origin in a sign change of
effective masses~\cite{lin85} instead of a band inversion.} where
$\Delta_0 < 0$, we consider also the modified \textit{Ansatz\/}
\begin{eqnarray}\label{eq:PankAns}
\Phi^{(n)}_j(r) &=& c_{n,j,\mathrm{a}}^{(\mathrm{P})}\,
\Phi^{(\mathrm{P})}_{j,\mathrm{a}}(r) + c_{n,j,
\mathrm{a}}^{(\mathrm{B})} \, \Phi^{(\mathrm{B})}_{j,\mathrm{a}}(r)
\nonumber \\[0.1cm] && \hspace{0.5cm} +\, c_{n,j,
\mathrm{b}}^{(\mathrm{P})}\, \Phi^{(\mathrm{P})}_{j,\mathrm{b}}(r) +
c_{n,j,\mathrm{b}}^{(\mathrm{B})} \, \Phi^{(\mathrm{B})}_{j,
\mathrm{b}}(r) \quad, \quad
\end{eqnarray}
which is obtained by replacing the propagating-wave-like part $c_{n,j,
\mathrm{a/b}}^{(\mathrm{D})}\,\Phi^{(\mathrm{D})}_{j,\mathrm{a/b}}(r)$
in Eq.~(\ref{eq:Ansatz}) by $c_{n,j,\mathrm{a/b}}^{(\mathrm{P})} \,
\Phi^{(\mathrm{P})}_{j,\mathrm{a/b}}(r)$ with the evanescent-wave
spinors~\footnote{Using identity 10.2.2 from Ref.~\cite{abr64},
it can be shown that the spinors in Eqs.~(\ref{eq:evanAns}) are
obtained from those in (\ref{eq:perpD}) by making the replacements $k
= i\bar k$ and $\gamma_k = i\, \bar\gamma_{\bar k}$.}
\begin{subequations}\label{eq:evanAns}
\begin{eqnarray}
\Phi^{(\mathrm{P})}_{j,\mathrm{a}}(r) &=& \frac{1}{\sqrt{\bar k\, r}}
\begin{pmatrix} I_j(\bar k\, r) \\[0.1cm] -i\,\bar\gamma_{\bar k}\,
I_{j+1}(\bar k\, r) \\[0.1cm] i\, I_j(\bar k\, r) \\[0.1cm]
-\bar\gamma_{\bar k} \, I_{j+1}(\bar k\, r) \end{pmatrix} \quad ,
\\[0.2cm] \Phi^{(\mathrm{P})}_{j,\mathrm{b}}(r) &=&
\frac{1}{\sqrt{\bar k\, r}} \begin{pmatrix} i\, I_{j+1}(\bar k\,
r)\\[0.1cm] \bar\gamma_{\bar k}\, I_j(\bar k\, r)\\[0.1cm] I_{j+1}
(\bar k\, r) \\[0.1cm] i\, \bar \gamma_{\bar k} \, I_j(\bar k\, r)
\end{pmatrix} \quad .
\end{eqnarray}
\end{subequations}
The energy-dependent parameters entering Eqs.~(\ref{eq:evanAns}) are
\begin{subequations}
\begin{eqnarray}
\tilde{\bar k} &=&  \left[ \frac{1}{2} \left( 1 + {\tilde\Delta}_0
- \sqrt{(1+{\tilde\Delta}_0)^2 + 4 {\tilde E}^2 - {\tilde
\Delta}_0^2}\right)\right]^{\frac{1}{2}} , \quad \\[0.2cm]
\bar\gamma_{\bar k} &=& \mathrm{sgn}(\Delta_0) \left[ \frac{\sqrt{(
1+{\tilde\Delta}_0)^2 + 4 \tilde E^2 - {\tilde\Delta}_0^2} - 1 - 2
\tilde E}{\sqrt{(1 + {\tilde\Delta}_0)^2 + 4 \tilde E^2 - {\tilde
\Delta}_0^2}-1+2\tilde E} \right]^{\frac{1}{2}} \,\, . \nonumber \\
\end{eqnarray}
\end{subequations}
Imposing the hard-wall boundary condition at $r=R$ yields the secular
equations
\begin{equation}\label{eq:secuEva}
- \bar\gamma_{\bar k}\, \bar\gamma_q = \left\{ \begin{array}{cl}
\frac{I_{j+1}(q R)}{I_j(q R)}\, \frac{I_j(\bar k R)}{I_{j+1}(\bar k
R)} & \quad \mbox{case a}, \\[0.2cm] \frac{I_j(q R)}{I_{j+1}(q R)}\,
\frac{I_{j+1}(\bar k R)}{I_j(\bar k R)} & \quad \mbox{case b}.
\end{array}\right.
\end{equation}
The coefficients in the \textit{Ansatz\/} (\ref{eq:PankAns}) involving
only evanescent eigenstates of ${\mathcal H}^\perp_{j+\frac{1}{2}}$
are
\begin{subequations}
\begin{eqnarray}
c_{n_\mathrm{a},j,\mathrm{a}}^{(\mathrm{P})} &=&
\bar{N}_{n_\mathrm{a},j} \, , \, c_{n_\mathrm{a},j,
\mathrm{a}}^{(\mathrm{B})} = \bar{N}_{n_\mathrm{a},j}\, \bar
\xi_{n_\mathrm{a},j,\mathrm{a}} \, , \nonumber \\ && \hspace{2.5cm} 
c_{n_\mathrm{a},j,\mathrm{b}}^{(\mathrm{P})} = c_{n_\mathrm{a},j,
\mathrm{b}}^{(\mathrm{B})} = 0 \,\, ,\\
c_{n_\mathrm{b},j,\mathrm{b}}^{(\mathrm{P})} &=&
\bar{N}_{n_\mathrm{b},j} \, , \, c_{n_\mathrm{b},j,
\mathrm{b}}^{(\mathrm{B})} = \bar{N}_{n_\mathrm{b},j}\, \bar
\xi_{n_\mathrm{b},j,\mathrm{b}} \, , \nonumber \\ && \hspace{2.5cm}
c_{n_\mathrm{b},j,\mathrm{a}}^{(\mathrm{P})} =
c_{n_\mathrm{b},j,\mathrm{a}}^{(\mathrm{B})} = 0 \,\, ,
\end{eqnarray}
with the parameters
\begin{eqnarray}
\bar\xi_{n_\mathrm{a},j,\mathrm{a}} &=& \left. \bar \gamma_{\bar k}\,
\sqrt{\frac{q}{\bar k}} \, \frac{I_{j+1}(\bar k R)}{I_{j+1}(q R)}
\right|_{E = E^{(n_\mathrm{a})}_j} \quad , \\[0.2cm]
\bar\xi_{n_\mathrm{b},j,\mathrm{b}} &=& -i\, \left. \bar \gamma_{\bar
k}\, \sqrt{\frac{q}{\bar k}} \, \frac{I_j(\bar k R)}{I_j(q R)}
\right|_{E = E^{(n_\mathrm{b})}_j} \quad . 
\end{eqnarray}
\end{subequations}

The general form (\ref{eq:Spinor}) of TI-nanoparticle bound-state wave
functions, together with the \textit{Ans\"atze\/} (\ref{eq:Ansatz})
and (\ref{eq:PankAns}) [using also (\ref{eq:perpD}), (\ref{eq:perpB}),
and (\ref{eq:evanAns})] implies that the electron eigenstates in TI
nanoparticles can be written as given in Eqs.~(\ref{eq:aState}) and
(\ref{eq:bState}). These expressions form the basis for deriving
Eq.~(\ref{eq:univWF}), which is one of the main results of this work.
We use the general convention where energy levels for fixed $j$ have
labels $n$ such that $E^{(n')}_j > E^{(n)}_j$ for $n'>n$ and $n>0$
($n<0$) if $E^{(n)}_j>0$ $\big( E^{(n)}_j<0\big)$. In addition to the
quantized energies among any fixed-$j$ series that arise as solutions
of the secular equations (\ref{eq:secu}), (\ref{eq:secuEva}) yields
two more bound-state levels with energies within the bulk-material
band gap when $\Delta_0$ is negative and $j$ below an $R$-dependent
critical value $j_\mathrm{c}$. For large $R$, $j_\mathrm{c}$ is also
large, and the many solutions of (\ref{eq:secuEva}) correspond to the
topologically protected surface states of a bulk-TI material. Reducing
$R$ successively eliminates such solutions for higher $j$. Eventually,
at $R = R_\mathrm{c}$, the energies of the $j=1/2$ sub-gap bound
states are pushed beyond $\pm|\Delta_0|/2$ and the TI-nanoparticle
spectrum becomes non-inverted. Figure~\ref{fig:topState} illustrates
this general behavior. Assuming $q R\gg 1$ (which is typically
satisfied) and $\Delta_0/(\gamma k_\Delta) > -1/2$ (which can be
violated in existing materials), we find the analytical expressions
\begin{subequations}
\begin{eqnarray}\label{eq:critiRad}
j_\mathrm{c} &=& \left\lfloor \sqrt{1 + \frac{\Delta_0}{\gamma
k_\Delta}}\, \frac{R}{R_0} -1 \right\rfloor \quad , \\[0.1cm]
R_\mathrm{c} &=& \frac{3}{2}\,\frac{R_0}{\sqrt{1 + \Delta_0/(\gamma
k_\Delta)}} \quad .
\end{eqnarray}
\end{subequations}

\section{Optical transitions between bound states in TI nanoparticles}
\label{sec:optProp}

Optical spectroscopy provides a useful tool to study the electronic
structure of semiconductor nanomaterials~\cite{hau04,pel12}, including
single nanoparticles~\cite{gus98} and their ensembles~\cite{fom98}. In
the following, we explore how the special properties of TI
nanoparticles are manifested in their optical spectra. We start by
discussing the general theoretical approach for treating confined
multi-band systems and then focus on our particular system of
interest.

Electron states in the TI nanoparticle have the form
\begin{equation}
\ket{\Psi^{(n)}_{j m \kappa}(\rr)} = \sum_{\tau \sigma} \big[
\Psi^{(n)}_{j m \kappa}(\rr)\big]_{\tau \sigma}\, \ket{\tau\,\sigma}
\quad ,
\end{equation}
where the $\ket{\tau\,\sigma}$ are the band-edge basis
functions~\cite{zha09,liu10,bre18} listed in Eq.~(\ref{eq:basis}), and
$\big[\Psi^{(n)}_{j m\kappa}(\rr)\big]_{\tau \sigma}$ denote entries
in the envelope-function spinor $\Psi^{(n)}_{j m \kappa}(\rr)$ given
in Eq.~(\ref{eq:univWF}). Optical transitions of electrons in this
system are mediated by the dipole matrix element, which is most
generally given as the sum of two contributions~\footnote{In our
notation, initial (final) states are labelled by quantum numbers $n,
j, m, \kappa$ ($n', j', m', \kappa'$). Equation~(\ref{eq:dipole}) is a
generalization of Eq.~(5.34) from Ref.~\cite{hau04} and, as the
latter, relies on the assumption that the envelope-function part of
electronic wave functions varies on much larger length scales than the
basis functions.};
\begin{equation}\label{eq:dipole}
\vek{d}^{n' j' m' \kappa'}_{n\, j\, m\, \kappa} = \vek{\mathcal D}^{n'
j' m' \kappa'}_{n\, j\, m\, \kappa} + \vek{\mathcal B}^{n' j' m'
\kappa'}_{n\, j\, m \, \kappa} \quad ,
\end{equation}
where
\begin{equation}\label{eq:DipEnv}
\vek{\mathcal D}^{n' j' m' \kappa'}_{n\, j\, m\, \kappa}=\sum_{\tau
\sigma}\, \, \int d^3 r \,\,\, \big[\Psi^{(n')}_{j' m' \kappa'}(\rr)
\big]_{\tau\sigma}^\ast \,\, e\,\rr \,\, \big[ \Psi^{(n)}_{j m \kappa}
(\rr) \big]_{\tau \sigma}
\end{equation}
is the dipole matrix element for the envelope part of the electronic
wave functions ($e$ is the electron charge), and
\begin{equation}
\vek{\mathcal B}^{n' j' m' \kappa'}_{n\, j\, m\, \kappa} = \sum_{\tau
\tau' \atop \sigma \sigma'} \mathcal{F}^{n' j' m' \kappa' \tau'
\sigma'}_{n\, j\, m \, \kappa\, \tau\, \sigma}\,\, \vek{d}^{\tau'
\sigma'}_{\tau\, \sigma}
\end{equation}
contains the contributions of dipole matrix elements $\vek{d}^{\tau'
\sigma'}_{\tau\, \sigma} \equiv \bra{\tau'\sigma'} e\,\rr\ket{\tau
\sigma}$ between band-edge basis states, which are renormalized by
the form factors
\begin{equation}
\mathcal{F}^{n' j' m' \kappa' \tau' \sigma'}_{n\, j\, m\, \kappa\,
\tau\, \sigma} = \int d^3 r \,\,\, \big[ \Psi^{(n')}_{j' m' \kappa'}
(\rr)\big]_{\tau' \sigma'}^\ast \,\, \big[ \Psi^{(n)}_{j m \kappa}
(\rr)\big]_{\tau \sigma}\quad .
\end{equation}

When optical properties of wide-band-gap semiconductor materials are
discussed, the basis-function-related contributions $\vek{\mathcal
B}^{n' j' m' \kappa'}_{n\, j\, m\, \kappa}$ are understood to
represent `interband' transitions~\cite{hau04} occurring between
states from the valence and conduction bands. In contrast, the
envelope-function contributions $\vek{\mathcal D}^{n' j' m'
\kappa'}_{n\, j\, m\, \kappa}$ in such systems are referred to as
`intraband' or `intersubband' because they involve initial and final
states that are size-quantized levels deriving from the same (either
conduction or valence) band. In our present case of interest,
these two types of contributions to the optical matrix element cannot
be neatly classified as either `interband' or `intraband' anymore, as
the amplitude of every possible transition will be influenced
significantly by both~\cite{pau13}.

The fundamental relationship~\cite{hau04} between optical matrix
elements for band-edge basis functions and linear-in-$\kk$ terms in
the $\kk\cdot\vek{p}$ Hamiltonian Eq.~(\ref{eq:3DbulkHam}) implies
\begin{equation}\label{eq:optKane}
\vek{d}^{\tau'\sigma'}_{\tau\, \sigma} = \tau\, \frac{i\, e}{\Delta_0}
\, \left( \left. \vek{\nabla}_\kk\, H \right|_{\kk=\vek{0}}
\right)_{\tau\, \sigma}^{\tau' \sigma'} \quad ,
\end{equation}
where we use the notation $(O)_{\tau\, \sigma}^{\tau'\sigma'}$ to
indicate matrix elements of an operator $O$ acting in $\kk\cdot
\vek{p}$ space. Using (\ref{eq:optKane}), we can conveniently rewrite
the basis-function contribution to the optical dipole matrix element
as
\begin{equation}\label{eq:kdotpDip}
\vek{\mathcal B}^{n' j' m' \kappa'}_{n\, j\, m\, \kappa} \equiv
\frac{e\, R_0}{2} \int d^3 r \,\, \big[ \Psi^{(n')}_{j' m' \kappa'}
(\rr) \big]^\dagger \, \tau_y\otimes \vek{\sigma}\,\, \Psi^{(n)}_{j m
\kappa}(\rr)\,\, .
\end{equation}
Together with (\ref{eq:DipEnv}), we can thus express the total dipole
matrix element for optical transitions entirely within
envelope-function space as shown in Eq.~(\ref{eq:dipMatGen}), with the
operator $\vek{d}$ given in Eq.~(\ref{eq:envDip}).

\begin{figure}[t]
\includegraphics[width=0.8\columnwidth]{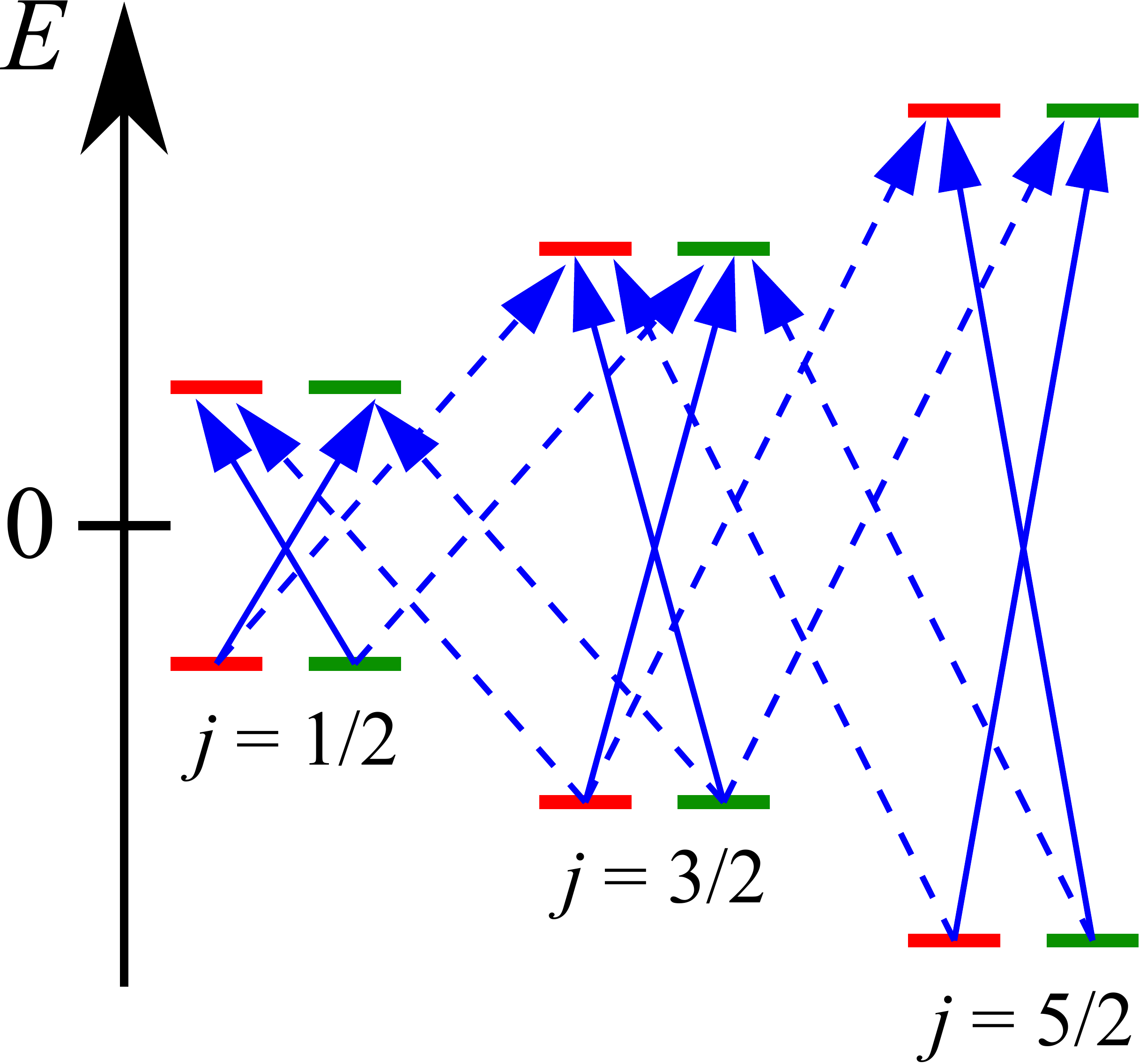}
\caption{\label{fig:trans}%
Optical transitions in undoped TI nanoparticles between levels derived
from topologically protected surface states. Red (green) color
indicates states with parity-related quantum number $\kappa = +$
($-$). Transitions conserving $\kappa$ (illustrated by dashed-line
arrows) require a change in the total-angular-momentum quantum number
$j$ by $\pm 1$. Transitions conserving $j$ (indicated by solid-line
arrows) are possible between states having opposite $\kappa$. The
selection rule $\Delta m = 0, \pm 1$ applies to all of these
transitions.}
\end{figure}

In the calculation of the envelope-function part (\ref{eq:DipEnv}) of
the optical-dipole matrix element (\ref{eq:dipole}), we employ the
standard~\cite{gre98} representation of real-space coordinates in
terms of spherical harmonics,
\begin{eqnarray}
x\pm i\, y &=& \mp\sqrt{\frac{8\pi}{3}}\, r \, Y_1^{\pm 1}(\theta,
\varphi) \quad , \\
z &=& \sqrt{\frac{4\pi}{3}}\, r \, Y_1^0(\theta, \varphi) \quad .
\end{eqnarray}
We also apply a well-known integral identity for spherical harmonics
[see, e.g., Eq.~(3.8.73) from Ref.~\cite{sak11}) to obtain
\begin{eqnarray}
&& \int_0^\pi d\theta\, \sin\theta \int_0^{2\pi} d\varphi \,\, \Big[
Y_{l^\prime}^{m^\prime}(\theta, \varphi) \big]^\ast Y_1^{m_1}(\theta,
\varphi)\, Y_l^m(\theta, \varphi) \nonumber \\[0.1cm] && =\,
\sqrt{\frac{3}{4\pi}}\, \sqrt{\frac{2 l + 1}{2 l^\prime + 1}}\,\,
\braket{l\, 0\, ; 1\, 0\, |\, l^\prime\, 0}\, \braket{l\, m\, ; 1\,
m_1\, |\, l^\prime\, m^\prime}\, , \quad
\end{eqnarray}
where the symbol $\braket{l_1\, m_1; l_2\, m_2\, |\, l \, m}$ is used
to denote Clebsch-Gordon coefficients~\cite{sak11}. Algebraic
simplification then yields the terms in Eqs.~(\ref{eq:transAmp}) that
are proportional to $\mathcal{R}_{n\, j}^{n' j'}$. Similarly, the
terms in (\ref{eq:transAmp}) that are proportional to the overlap
integrals $\mathcal{S}_{n\, j, \tau}^{n' j', \tau'}$ are obtained from
calculating the basis-function contribution (\ref{eq:kdotpDip}) to the
optical-dipole matrix element (\ref{eq:dipole}) using the
orthogonality relation for spherical harmonics.

\begin{figure}[t]
\includegraphics[width=0.95\columnwidth]{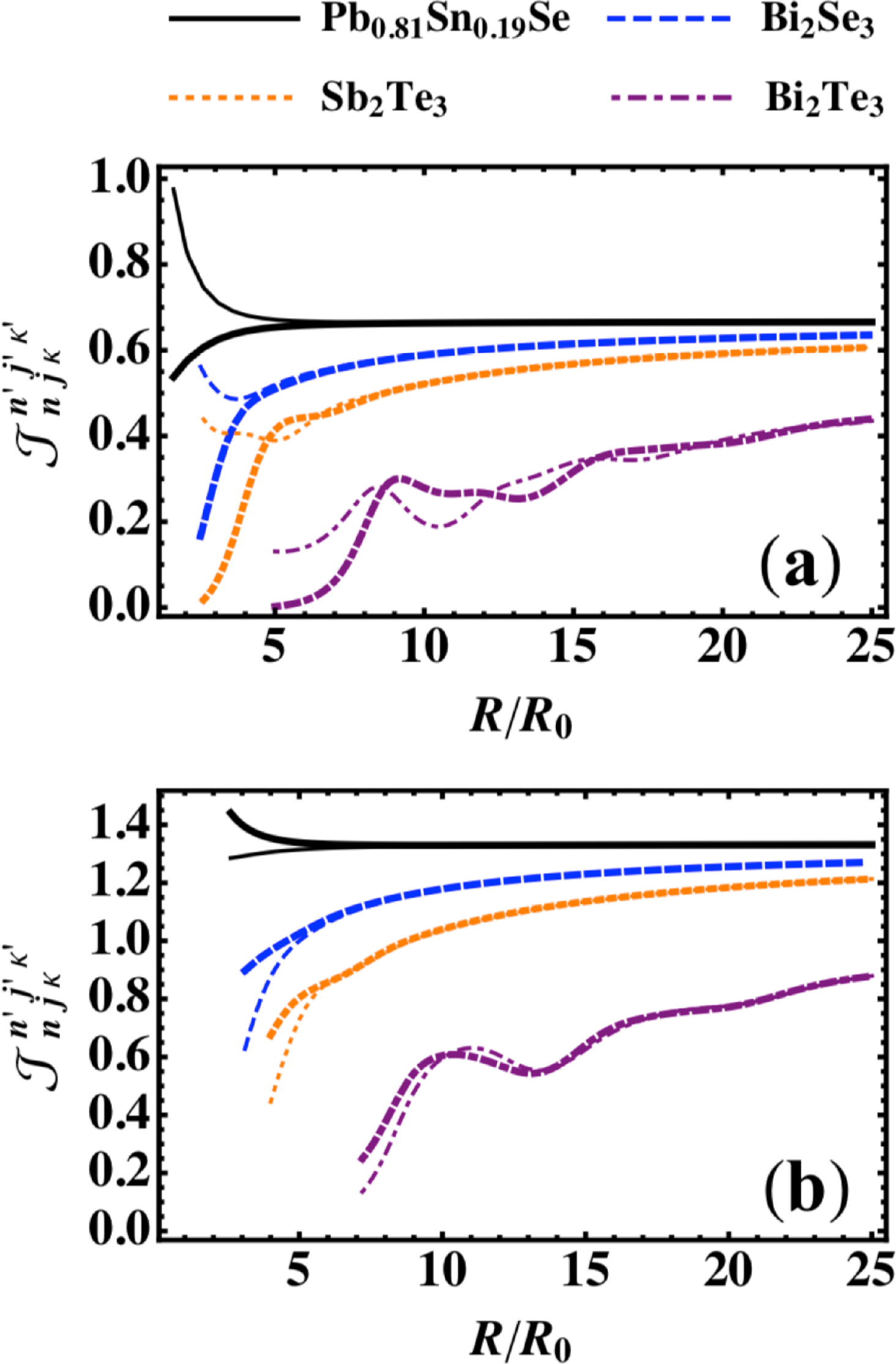}
\caption{\label{fig:probVsR}%
The modulus squared of the dipole moment for optical transitions in a
spherical topological-insulator nanoparticle with radius $R$ is given
by $e^2 R^2\, \mathcal{T}^{n' j' \kappa'}_{n\, j\, \kappa}$. Panel~(a)
plots the nanoparticle-size dependence of the dimensionless dipole
magnitude $\mathcal{T}^{n' j' \kappa'}_{n\, j\, \kappa}$ for the
lowest-energy transitions in an undoped nanoparticle, i.e., for
$n=-1$, $n'=1$, $j=j'=1/2$, and $\kappa = -\kappa' = +/-$
(thick/thin curves). Panel~(b) shows $\mathcal{T}^{n' j' \kappa'}_{n\,
j\, \kappa}$ for the next-highest-energy transitions where $n=-1$,
$n'=1$, $j=j'-1=1/2$, and $\kappa = \kappa' = +/-$ (thick/thin
curves). Here $R_0 = 2\gamma/|\Delta_0|$ is the bulk Compton length,
and the black solid (blue dashed, orange dotted, purple dot-dashed)
curve is obtained for $\Delta_0/(\gamma k_\Delta) = -0.039$ ($-1.2$,
$-2.3$, $-9.5$,), which is representative of Pb$_{0.81}$Sn$_{0.19}$Se
(Bi$_2$Se$_3$, Sb$_2$Te$_3$, Bi$_2$Te$_3$).}
\end{figure}

Results from Sec.~\ref{sec:states} imply that in-gap bound states from
the case-a series have eigenenergies with equal magnitude but opposite
sign as those of the case-b series. Each of these levels is degenerate
in the parity-related quantum number $\kappa$. Hence, in an undoped
TI nanoparticle, optical transitions can occur between the
opposite-energy eigenstates, and the energy cutoff for photon
absorption and emission via single-electron transitions is given by
the TI-nanoparticle band gap $2\cdot \mathrm{min}\big\{ E_j^{(1)}
\big\}$. We illustrate the allowed optical transitions between
topological (i.e., in-gap) nanoparticle bound states in
Fig.~\ref{fig:trans}. Such single-particle transitions will dominantly
determine the optical properties of nanoparticles with sizes smaller
than the charge-carriers' Bohr radius~\cite{efr82}. The total
magnitude of optical transition matrix elements associated with such
transitions is given by
\begin{subequations}
\begin{eqnarray}
\left( T_\pm \right)^{n' j' \kappa'}_{n\, j\, \kappa} &=& \sum_{m=
-j}^j \Big| (d_x \pm i d_y)^{n' j' m' \kappa'}_{n\, j\, m\, \kappa}
\Big|^2 \equiv \frac{2 e^2 R^2}{3} \, \mathcal{T}^{n' j' \kappa'}_{n\,
j\, \kappa} , \nonumber \\[-0.3cm] \\
\left( T_z \right)^{n' j' \kappa'}_{n\, j\, \kappa} &=& \sum_{m=-j}^j
\Big| (d_z)^{n' j' m' \kappa'}_{n\, j\, m\, \kappa} \Big|^2 \equiv
\frac{e^2 R^2}{3} \, \mathcal{T}^{n' j' \kappa'}_{n\, j\, \kappa}
\,\, ,
\end{eqnarray}
\end{subequations}
with a dimensionless measure of the squared magnitude of the optical
dipole vector defined as
\begin{widetext}
\begin{eqnarray}
\mathcal{T}^{n' j' \kappa'}_{n\, j\, \kappa} &=& \delta_{\kappa',
\kappa} \sum_{\xi = \pm 1} \frac{(2j+\xi)[2(j+1)+\xi]}{2(2j+1+\xi)}
\left| \frac{\mathcal{R}_{n\, j}^{n' j'}}{R} - i \kappa \xi \,
\frac{R_0}{2 R}\,\, \mathcal{S}_{n\, j, -\kappa\xi}^{n' j', \kappa\xi}
\right|^2 \delta_{j', j+\xi} \nonumber \\[0.1cm] && \hspace{3cm} +\,\,
\delta_{\kappa', -\kappa} \,\, \frac{2j+1}{4 j (j+1)} \left|
\frac{\mathcal{R}_{n\, j}^{n' j'}}{R} + i\kappa\, \frac{R_0}{2 R}
\left( j \, \mathcal{S}_{n \, j, -\kappa}^{n' j', \kappa} + (j+1)\,
\mathcal{S}_{n \, j, \kappa}^{n' j', -\kappa} \right) \right|^2
\delta_{j', j} \quad .
\end{eqnarray}
\end{widetext}
The nanoparticle-size dependence of $\mathcal{T}^{n' j' \kappa'}_{n\,
j\, \kappa}$ for the two lowest-energy transitions is plotted in
Fig.~\ref{fig:probVsR}. It exhibits a confinement-induced
parity-symmetry breaking that gets most pronounced in the small-size
limit where transition amplitudes are dominated by overlap integrals
$\mathcal{S}_{n \, j, \pm}^{n' j', \mp}$ whose magnitude deviates from
unity.

\section{Conclusions and outlook}
\label{sec:outlook}

We have solved analytically the quantum-confinement problem of charge
carriers in spherical nanoparticles made from topological-insulator
(TI) materials, extending previous results~\cite{imu12,pau13}. The
mathematical formalism employed in our study unifies related
multi-band descriptions of semiconductor nanocrystals~\cite{she67,
efr98} and confined Dirac fermions~\cite{alb96,abr02,lay11,pau13}.
Compact transcendental secular equations are obtained for the energies
of both ordinary (i.e., deriving from linear combinations of
propagating bulk states) and topological (i.e., constituting confined
intrinsic surface states) bound-state levels. See Eqs.~(\ref{eq:secu})
and (\ref{eq:secuEva}), respectively. Explicit results for the size
dependence of topological-bound-state sub-gap energies calculated
using parameters applicable to real materials are shown in
Fig.~\ref{fig:topState}. Significant deviations from previously
derived~\cite{imu12,pau13} asymptotic expressions arise already for
not-so-small values of $R/R_0$, where $R$ and $R_0$ are the
nanoparticle radius and the bulk-material Compton length,
respectively.

A universal form of the envelope-wave-function spinors for
TI-nanoparticle bound states in terms of spherical harmonics is
presented in Eq.~(\ref{eq:univWF}). These states are labeled by the
half-integer quantum numbers for total angular momentum ($j$) and its
projection onto an arbitrary axis ($m$), as well as a two-valued
quantum number $\kappa=\pm 1$. States with fixed $j$, $m$ and $\kappa$
are also eigenstates of parity with eigenvalue $(-1)^{j-
\frac{\kappa}{2}}$.

We considered optical transitions between single-electron bound states
in a TI nanoparticle via the conventional~\cite{hau04} electric-dipole
coupling. Our approach describes both envelope-function- and
basis-state-mediated transitions via a compact optical-dipole operator
in envelope-function space [see Eqs.~(\ref{eq:envDip}) and
(\ref{eq:dipMatGen})], and general analytical results are obtained for
the transition matrix elements [Eqs.~(\ref{eq:transAmp})]. In addition
to the previously discussed~\cite{pau13} transitions between levels
where $j'-j=\pm 1$, we also identify possible transitions where $j'-j
=0$. The latter were not identified in Ref.~\cite{pau13}. The possibility
to conserve total angular momentum in an optical transition arises
because there exist TI-nanoparticle bound states at any given $j$ that
have opposite parity. For such a pair of states [see their explicit
expressions in Eq.~(\ref{eq:univWF})], corresponding spinor entries
have orbital angular momentum differing by one unit, as required for
accommodating a photon's angular momentum. The multitude of possible
transitions between topological bound-state levels in an undoped TI
nanoparticle is illustrated in Fig.~\ref{fig:trans}.

The effect of the nanoparticle size on optical transitions is
contained in the radial-wave-function overlap integrals given in
Eqs.~(\ref{eq:effectRad}) and (\ref{eq:radOlap}). Their size
dependence for the lowest subgap-energy bound state is plotted in
Fig.~\ref{fig:radMat}. For $R\gg R_0$, the envelope-function
contribution to the transition matrix element is $\propto R$,
reflecting the surface-localized character of these states in this
limit. However, significant deviations from this expected behavior
occur already from $R\sim 10\, R_0$ for typical materials, and also
the basis-state contribution starts to become relevant within this
size range. The nanoparticle-radius dependence of transition-dipole
magnitudes for the lowest-energy unconventional and conventional types
of optical transitions is shown in Fig.~\ref{fig:probVsR}. Both have
similar order of magnitude and exhibit pronounced parity-symmetry
breaking in the small-$R$ limit.

Our results provide detailed information about how size quantization
affects surface-state-related electronic and optical properties in TI
nanoparticles, which has important practical implications. To start with,
the existence of additional optical transitions beyond those identified
earlier in Ref.~\cite{pau13} requires re-evaluation of suggested
schemes~\cite{pau13} to employ TI-nanoparticle states in
quantum information processing. More generally, our theoretical
description can serve as a starting point for developing any type of
application that relies on specifics of the electronic energy spectrum
and quantum transitions between TI-nanoparticle bound states. For
example, some of the currently envisioned uses of TI nanoparticles in
optoelectronic~\cite{pol17,rid19} and thermoelectric~\cite{xu17} devices
are contingent on the presence of topologically protected surface states.
Recognizing the existence, and strongly materials-dependent magnitude,
of a critical radius below which the topological surface states disappear
is therefore of crucial importance. The emergent gaplessness of the
Bi$_2$Te$_3$-nanopoarticle spectrum for certain sizes [see
Fig.~\ref{fig:topState}(b)] could become a platform for intriguing
applications, including the possibility to host unconventional Majorana
excitations as was proposed for gapless TI nanoribbons~\cite{man17}.

Our theoretical approach is based on the envelope-function
continuum-model description for the TI band structure~\cite{zha09,
liu10,bre18}. The mathematical structure of this formalism
necessitates inclusion of fast-decaying wave-function contributions
[those given in Eqs.~(\ref{eq:perpB})] representing the effect of
remote bands  whose influence on observable physical quantities needs
to be interpreted with caution~\cite{whi81,sch85,kli18}. Typically,
the physically significant wave functions from Eqs.~(\ref{eq:perpD})
and (\ref{eq:evanAns}) dominate the system's low-energy properties.
However, in confined TI materials with large values of $|\Delta_0|/
(\gamma k_\Delta)$, we find that both types of evanescent states
[(\ref{eq:evanAns}) and (\ref{eq:perpB})] present in the sub-gap
bound-state wave functions (\ref{eq:PankAns}) can be relevant on
comparable length scales, signaling the impending breakdown of the
continuum-model approach. Nevertheless, qualitative trends exhibited
in size-dependent physical properties are expected to be accurate even
relatively close to this extreme limit. Thus our approach serves as a
useful tool for modeling the effect of quantum confinement in TI
nanoparticles, especially as long as atomistic band-structure
calculations~\cite{ban16} can be performed only for very small
systems~\cite{var14,nec16}. Results presented here also provide an
interesting reference point for understanding size-quantization
effects in other TI nanostructures such as cylindrical quantum
dots~\cite{her14} or nanowires with finite~\cite{kun11} and infinite
length~\cite{imu11,ior16,vir18}.

In future work, the obtained single-particle wave functions for
TI-nanoparticle bound states could serve as input to discuss many-body
effects due to Coulomb interactions. In particular, exploring the
properties of excitons would be of interest in the context of optical
properties of TI nanoparticles~\cite{mal10}. The multi-band spinor
structure of charge carriers' quantum states should result in a
form-factor renormalization of the Coulomb-potential matrix element
similar to previously considered situations~\cite{ker13}. Interesting
properties associated with two-particle states of Dirac-like
carriers~\cite{sab10,ber13}, e.g., the recently predicted existence of
an internal angular momentum of the exciton~\cite{tru18}, could be
elucidated in greater detail. Early numerical studies~\cite{mal10} can
serve as a benchmark for evaluating the accuracy of approximate
analytical results.

\begin{acknowledgments}

The authors benefited from useful discussions with K.-I.~Imura. The
work of M.G.C.\ was partially supported by a Victoria University of
Wellington Summer Research Scholarship. L.G.\ is the grateful
recipient of a Vanier Canada Graduate Scholarship. Research at
Perimeter Institute is supported in part by the Government of
Canada through the Department of Innovation, Science and Economic
Development Canada and by the Province of Ontario through the
Ministry of Economic Development, Job Creation and Trade.

\end{acknowledgments}


\begin{thebibliography}{83}%
\makeatletter
\providecommand \@ifxundefined [1]{%
 \@ifx{#1\undefined}
}%
\providecommand \@ifnum [1]{%
 \ifnum #1\expandafter \@firstoftwo
 \else \expandafter \@secondoftwo
 \fi
}%
\providecommand \@ifx [1]{%
 \ifx #1\expandafter \@firstoftwo
 \else \expandafter \@secondoftwo
 \fi
}%
\providecommand \natexlab [1]{#1}%
\providecommand \enquote  [1]{``#1''}%
\providecommand \bibnamefont  [1]{#1}%
\providecommand \bibfnamefont [1]{#1}%
\providecommand \citenamefont [1]{#1}%
\providecommand \href@noop [0]{\@secondoftwo}%
\providecommand \href [0]{\begingroup \@sanitize@url \@href}%
\providecommand \@href[1]{\@@startlink{#1}\@@href}%
\providecommand \@@href[1]{\endgroup#1\@@endlink}%
\providecommand \@sanitize@url [0]{\catcode `\\12\catcode `\$12\catcode
  `\&12\catcode `\#12\catcode `\^12\catcode `\_12\catcode `\%12\relax}%
\providecommand \@@startlink[1]{}%
\providecommand \@@endlink[0]{}%
\providecommand \url  [0]{\begingroup\@sanitize@url \@url }%
\providecommand \@url [1]{\endgroup\@href {#1}{\urlprefix }}%
\providecommand \urlprefix  [0]{URL }%
\providecommand \Eprint [0]{\href }%
\providecommand \doibase [0]{https://doi.org/}%
\providecommand \selectlanguage [0]{\@gobble}%
\providecommand \bibinfo  [0]{\@secondoftwo}%
\providecommand \bibfield  [0]{\@secondoftwo}%
\providecommand \translation [1]{[#1]}%
\providecommand \BibitemOpen [0]{}%
\providecommand \bibitemStop [0]{}%
\providecommand \bibitemNoStop [0]{.\EOS\space}%
\providecommand \EOS [0]{\spacefactor3000\relax}%
\providecommand \BibitemShut  [1]{\csname bibitem#1\endcsname}%
\let\auto@bib@innerbib\@empty
\bibitem [{\citenamefont {Yoffe}(1993)}]{yof93}%
  \BibitemOpen
  \bibfield  {author} {\bibinfo {author} {\bibfnamefont {A.~D.}\ \bibnamefont
  {Yoffe}},\ }\bibfield  {title} {\bibinfo {title} {Low-dimensional systems:
  quantum size effects and electronic properties of semiconductor
  microcrystallites (zero-dimensional systems) and some quasi-two-dimensional
  systems},\ }\href {https://doi.org/10.1080/00018739300101484} {\bibfield
  {journal} {\bibinfo  {journal} {Adv. Phys.}\ }\textbf {\bibinfo {volume}
  {42}},\ \bibinfo {pages} {173} (\bibinfo {year} {1993})}\BibitemShut
  {NoStop}%
\bibitem [{\citenamefont {Efros}\ and\ \citenamefont {Rosen}(2000)}]{efr00}%
  \BibitemOpen
  \bibfield  {author} {\bibinfo {author} {\bibfnamefont {A.~L.}\ \bibnamefont
  {Efros}}\ and\ \bibinfo {author} {\bibfnamefont {M.}~\bibnamefont {Rosen}},\
  }\bibfield  {title} {\bibinfo {title} {The electronic structure of
  semiconductor nanocrystals},\ }\href
  {https://doi.org/10.1146/annurev.matsci.30.1.475} {\bibfield  {journal}
  {\bibinfo  {journal} {Annu. Rev. Mater. Sci.}\ }\textbf {\bibinfo {volume}
  {30}},\ \bibinfo {pages} {475} (\bibinfo {year} {2000})}\BibitemShut
  {NoStop}%
\bibitem [{\citenamefont {Cho}\ \emph {et~al.}(2012)\citenamefont {Cho},
  \citenamefont {Kim}, \citenamefont {Syers}, \citenamefont {Butch},
  \citenamefont {Paglione},\ and\ \citenamefont {Fuhrer}}]{cho12}%
  \BibitemOpen
  \bibfield  {author} {\bibinfo {author} {\bibfnamefont {S.}~\bibnamefont
  {Cho}}, \bibinfo {author} {\bibfnamefont {D.}~\bibnamefont {Kim}}, \bibinfo
  {author} {\bibfnamefont {P.}~\bibnamefont {Syers}}, \bibinfo {author}
  {\bibfnamefont {N.~P.}\ \bibnamefont {Butch}}, \bibinfo {author}
  {\bibfnamefont {J.}~\bibnamefont {Paglione}},\ and\ \bibinfo {author}
  {\bibfnamefont {M.~S.}\ \bibnamefont {Fuhrer}},\ }\bibfield  {title}
  {\bibinfo {title} {Topological insulator quantum dot with tunable barriers},\
  }\href {https://doi.org/10.1021/nl203851g} {\bibfield  {journal} {\bibinfo
  {journal} {Nano Lett.}\ }\textbf {\bibinfo {volume} {12}},\ \bibinfo {pages}
  {469} (\bibinfo {year} {2012})}\BibitemShut {NoStop}%
\bibitem [{\citenamefont {Kershaw}\ \emph {et~al.}(2013)\citenamefont
  {Kershaw}, \citenamefont {Susha},\ and\ \citenamefont {Rogach}}]{ker13a}%
  \BibitemOpen
  \bibfield  {author} {\bibinfo {author} {\bibfnamefont {S.~V.}\ \bibnamefont
  {Kershaw}}, \bibinfo {author} {\bibfnamefont {A.~S.}\ \bibnamefont {Susha}},\
  and\ \bibinfo {author} {\bibfnamefont {A.~L.}\ \bibnamefont {Rogach}},\
  }\bibfield  {title} {\bibinfo {title} {Narrow bandgap colloidal metal
  chalcogenide quantum dots: synthetic methods, heterostructures, assemblies,
  electronic and infrared optical properties},\ }\href
  {https://doi.org/10.1039/C2CS35331H} {\bibfield  {journal} {\bibinfo
  {journal} {Chem. Soc. Rev.}\ }\textbf {\bibinfo {volume} {42}},\ \bibinfo
  {pages} {3033} (\bibinfo {year} {2013})}\BibitemShut {NoStop}%
\bibitem [{\citenamefont {Vargas}\ \emph {et~al.}(2014)\citenamefont {Vargas},
  \citenamefont {Basak}, \citenamefont {Liu}, \citenamefont {Wang},
  \citenamefont {Panaitescu}, \citenamefont {Lin}, \citenamefont {Markiewicz},
  \citenamefont {Bansil},\ and\ \citenamefont {Kar}}]{var14}%
  \BibitemOpen
  \bibfield  {author} {\bibinfo {author} {\bibfnamefont {A.}~\bibnamefont
  {Vargas}}, \bibinfo {author} {\bibfnamefont {S.}~\bibnamefont {Basak}},
  \bibinfo {author} {\bibfnamefont {F.}~\bibnamefont {Liu}}, \bibinfo {author}
  {\bibfnamefont {B.}~\bibnamefont {Wang}}, \bibinfo {author} {\bibfnamefont
  {E.}~\bibnamefont {Panaitescu}}, \bibinfo {author} {\bibfnamefont
  {H.}~\bibnamefont {Lin}}, \bibinfo {author} {\bibfnamefont {R.}~\bibnamefont
  {Markiewicz}}, \bibinfo {author} {\bibfnamefont {A.}~\bibnamefont {Bansil}},\
  and\ \bibinfo {author} {\bibfnamefont {S.}~\bibnamefont {Kar}},\ }\bibfield
  {title} {\bibinfo {title} {The changing colors of a quantum-confined
  topological insulator},\ }\href {https://doi.org/10.1021/nn404013d}
  {\bibfield  {journal} {\bibinfo  {journal} {ACS Nano}\ }\textbf {\bibinfo
  {volume} {8}},\ \bibinfo {pages} {1222} (\bibinfo {year} {2014})}\BibitemShut
  {NoStop}%
\bibitem [{\citenamefont {Jia}\ \emph {et~al.}(2015)\citenamefont {Jia},
  \citenamefont {Lou}, \citenamefont {Cheng}, \citenamefont {Wang},
  \citenamefont {Yao}, \citenamefont {Dai}, \citenamefont {Lin},\ and\
  \citenamefont {Chang}}]{jia15}%
  \BibitemOpen
  \bibfield  {author} {\bibinfo {author} {\bibfnamefont {G.~Z.}\ \bibnamefont
  {Jia}}, \bibinfo {author} {\bibfnamefont {W.~K.}\ \bibnamefont {Lou}},
  \bibinfo {author} {\bibfnamefont {F.}~\bibnamefont {Cheng}}, \bibinfo
  {author} {\bibfnamefont {X.~L.}\ \bibnamefont {Wang}}, \bibinfo {author}
  {\bibfnamefont {J.~H.}\ \bibnamefont {Yao}}, \bibinfo {author} {\bibfnamefont
  {N.}~\bibnamefont {Dai}}, \bibinfo {author} {\bibfnamefont {H.~Q.}\
  \bibnamefont {Lin}},\ and\ \bibinfo {author} {\bibfnamefont {K.}~\bibnamefont
  {Chang}},\ }\bibfield  {title} {\bibinfo {title} {Excellent photothermal
  conversion of core/shell {CdSe/Bi2Se3} quantum dots},\ }\href
  {https://doi.org/10.1007/s12274-014-0629-2} {\bibfield  {journal} {\bibinfo
  {journal} {Nano Res.}\ }\textbf {\bibinfo {volume} {8}},\ \bibinfo {pages}
  {1443} (\bibinfo {year} {2015})}\BibitemShut {NoStop}%
\bibitem [{\citenamefont {Vargas}\ \emph {et~al.}(2015)\citenamefont {Vargas},
  \citenamefont {Liu},\ and\ \citenamefont {Kar}}]{var15}%
  \BibitemOpen
  \bibfield  {author} {\bibinfo {author} {\bibfnamefont {A.}~\bibnamefont
  {Vargas}}, \bibinfo {author} {\bibfnamefont {F.}~\bibnamefont {Liu}},\ and\
  \bibinfo {author} {\bibfnamefont {S.}~\bibnamefont {Kar}},\ }\bibfield
  {title} {\bibinfo {title} {Giant enhancement of light emission from nanoscale
  {Bi2Se3}},\ }\href {https://doi.org/10.1063/1.4922729} {\bibfield  {journal}
  {\bibinfo  {journal} {Appl. Phys. Lett.}\ }\textbf {\bibinfo {volume}
  {106}},\ \bibinfo {pages} {243107} (\bibinfo {year} {2015})}\BibitemShut
  {NoStop}%
\bibitem [{\citenamefont {Claro}\ \emph {et~al.}(2019)\citenamefont {Claro},
  \citenamefont {Levy}, \citenamefont {Gangopadhyay}, \citenamefont {Smith},\
  and\ \citenamefont {Tamargo}}]{cla19}%
  \BibitemOpen
  \bibfield  {author} {\bibinfo {author} {\bibfnamefont {M.~S.}\ \bibnamefont
  {Claro}}, \bibinfo {author} {\bibfnamefont {I.}~\bibnamefont {Levy}},
  \bibinfo {author} {\bibfnamefont {A.}~\bibnamefont {Gangopadhyay}}, \bibinfo
  {author} {\bibfnamefont {D.~J.}\ \bibnamefont {Smith}},\ and\ \bibinfo
  {author} {\bibfnamefont {M.~C.}\ \bibnamefont {Tamargo}},\ }\bibfield
  {title} {\bibinfo {title} {Self-assembled bismuth selenide ({Bi2Se3}) quantum
  dots grown by molecular beam epitaxy},\ }\href
  {https://doi.org/10.1038/s41598-019-39821-y} {\bibfield  {journal} {\bibinfo
  {journal} {Sci. Rep.}\ }\textbf {\bibinfo {volume} {9}},\ \bibinfo {pages}
  {3370} (\bibinfo {year} {2019})}\BibitemShut {NoStop}%
\bibitem [{\citenamefont {S.Rider}\ \emph {et~al.}(2019)\citenamefont
  {S.Rider}, \citenamefont {Sokolikova}, \citenamefont {Hanham}, \citenamefont
  {Navarro-C{\'\i}a}, \citenamefont {Haynes}, \citenamefont {Lee},
  \citenamefont {Daniele}, \citenamefont {Guidi}, \citenamefont {Mattevi},
  \citenamefont {Lupi},\ and\ \citenamefont {Giannini}}]{rid19}%
  \BibitemOpen
  \bibfield  {author} {\bibinfo {author} {\bibfnamefont {M.}~\bibnamefont
  {S.Rider}}, \bibinfo {author} {\bibfnamefont {M.}~\bibnamefont {Sokolikova}},
  \bibinfo {author} {\bibfnamefont {S.~M.}\ \bibnamefont {Hanham}}, \bibinfo
  {author} {\bibfnamefont {M.}~\bibnamefont {Navarro-C{\'\i}a}}, \bibinfo
  {author} {\bibfnamefont {P.}~\bibnamefont {Haynes}}, \bibinfo {author}
  {\bibfnamefont {D.}~\bibnamefont {Lee}}, \bibinfo {author} {\bibfnamefont
  {M.}~\bibnamefont {Daniele}}, \bibinfo {author} {\bibfnamefont {M.~C.}\
  \bibnamefont {Guidi}}, \bibinfo {author} {\bibfnamefont {C.}~\bibnamefont
  {Mattevi}}, \bibinfo {author} {\bibfnamefont {S.}~\bibnamefont {Lupi}},\ and\
  \bibinfo {author} {\bibfnamefont {V.}~\bibnamefont {Giannini}},\ }\bibfield
  {title} {\bibinfo {title} {Experimental signature of a topological quantum
  dot},\ }\Eprint {https://arxiv.org/abs/1905.06193} {arXiv:1905.06193
  [cond-mat.mes-hall]}  (\bibinfo {year} {2019})\BibitemShut {NoStop}%
\bibitem [{\citenamefont {Hong}\ \emph {et~al.}(2014)\citenamefont {Hong},
  \citenamefont {Kong},\ and\ \citenamefont {Cui}}]{hon14}%
  \BibitemOpen
  \bibfield  {author} {\bibinfo {author} {\bibfnamefont {S.~S.}\ \bibnamefont
  {Hong}}, \bibinfo {author} {\bibfnamefont {D.}~\bibnamefont {Kong}},\ and\
  \bibinfo {author} {\bibfnamefont {Y.}~\bibnamefont {Cui}},\ }\bibfield
  {title} {\bibinfo {title} {Topological insulator nanostructures},\ }\href
  {https://doi.org/10.1557/mrs.2014.196} {\bibfield  {journal} {\bibinfo
  {journal} {MRS Bull.}\ }\textbf {\bibinfo {volume} {39}},\ \bibinfo {pages}
  {873} (\bibinfo {year} {2014})}\BibitemShut {NoStop}%
\bibitem [{\citenamefont {Hasan}\ and\ \citenamefont {Kane}(2010)}]{has10}%
  \BibitemOpen
  \bibfield  {author} {\bibinfo {author} {\bibfnamefont {M.~Z.}\ \bibnamefont
  {Hasan}}\ and\ \bibinfo {author} {\bibfnamefont {C.~L.}\ \bibnamefont
  {Kane}},\ }\bibfield  {title} {\bibinfo {title} {\textit{Colloquium}:
  Topological insulators},\ }\href {https://doi.org/10.1103/RevModPhys.82.3045}
  {\bibfield  {journal} {\bibinfo  {journal} {Rev. Mod. Phys.}\ }\textbf
  {\bibinfo {volume} {82}},\ \bibinfo {pages} {3045} (\bibinfo {year}
  {2010})}\BibitemShut {NoStop}%
\bibitem [{\citenamefont {Hasan}\ and\ \citenamefont {Moore}(2011)}]{has11}%
  \BibitemOpen
  \bibfield  {author} {\bibinfo {author} {\bibfnamefont {M.~Z.}\ \bibnamefont
  {Hasan}}\ and\ \bibinfo {author} {\bibfnamefont {J.~E.}\ \bibnamefont
  {Moore}},\ }\bibfield  {title} {\bibinfo {title} {Three-dimensional
  topological insulators},\ }\href
  {https://doi.org/10.1146/annurev-conmatphys-062910-140432} {\bibfield
  {journal} {\bibinfo  {journal} {Annu. Rev. Cond. Mat. Phys.}\ }\textbf
  {\bibinfo {volume} {2}},\ \bibinfo {pages} {55} (\bibinfo {year}
  {2011})}\BibitemShut {NoStop}%
\bibitem [{\citenamefont {Hasan}\ \emph {et~al.}(2015)\citenamefont {Hasan},
  \citenamefont {Xu},\ and\ \citenamefont {Bian}}]{has15}%
  \BibitemOpen
  \bibfield  {author} {\bibinfo {author} {\bibfnamefont {M.~Z.}\ \bibnamefont
  {Hasan}}, \bibinfo {author} {\bibfnamefont {S.-Y.}\ \bibnamefont {Xu}},\ and\
  \bibinfo {author} {\bibfnamefont {G.}~\bibnamefont {Bian}},\ }\bibfield
  {title} {\bibinfo {title} {Topological insulators, topological
  superconductors and {W}eyl fermion semimetals: discoveries, perspectives and
  outlooks},\ }\href {https://doi.org/10.1088/0031-8949/2015/t164/014001}
  {\bibfield  {journal} {\bibinfo  {journal} {Phys. Scr.}\ }\textbf {\bibinfo
  {volume} {T164}},\ \bibinfo {pages} {014001} (\bibinfo {year}
  {2015})}\BibitemShut {NoStop}%
\bibitem [{\citenamefont {Politano}\ \emph {et~al.}(2017)\citenamefont
  {Politano}, \citenamefont {Viti},\ and\ \citenamefont {Vitiello}}]{pol17}%
  \BibitemOpen
  \bibfield  {author} {\bibinfo {author} {\bibfnamefont {A.}~\bibnamefont
  {Politano}}, \bibinfo {author} {\bibfnamefont {L.}~\bibnamefont {Viti}},\
  and\ \bibinfo {author} {\bibfnamefont {M.~S.}\ \bibnamefont {Vitiello}},\
  }\bibfield  {title} {\bibinfo {title} {Optoelectronic devices, plasmonics,
  and photonics with topological insulators},\ }\href
  {https://doi.org/10.1063/1.4977782} {\bibfield  {journal} {\bibinfo
  {journal} {APL Mater.}\ }\textbf {\bibinfo {volume} {5}},\ \bibinfo {pages}
  {035504} (\bibinfo {year} {2017})}\BibitemShut {NoStop}%
\bibitem [{\citenamefont {McIver}\ \emph {et~al.}(2012)\citenamefont {McIver},
  \citenamefont {Hsieh}, \citenamefont {Steinberg}, \citenamefont
  {Jarillo-Herrero},\ and\ \citenamefont {Gedik}}]{mci12}%
  \BibitemOpen
  \bibfield  {author} {\bibinfo {author} {\bibfnamefont {J.~W.}\ \bibnamefont
  {McIver}}, \bibinfo {author} {\bibfnamefont {D.}~\bibnamefont {Hsieh}},
  \bibinfo {author} {\bibfnamefont {H.}~\bibnamefont {Steinberg}}, \bibinfo
  {author} {\bibfnamefont {P.}~\bibnamefont {Jarillo-Herrero}},\ and\ \bibinfo
  {author} {\bibfnamefont {N.}~\bibnamefont {Gedik}},\ }\bibfield  {title}
  {\bibinfo {title} {Control over topological insulator photocurrents with
  light polarization},\ }\href {https://doi.org/10.1038/NNANO.2011.214}
  {\bibfield  {journal} {\bibinfo  {journal} {Nat. Nanotech.}\ }\textbf
  {\bibinfo {volume} {7}},\ \bibinfo {pages} {96} (\bibinfo {year}
  {2012})}\BibitemShut {NoStop}%
\bibitem [{\citenamefont {Tian}\ \emph {et~al.}(2017)\citenamefont {Tian},
  \citenamefont {Hong}, \citenamefont {Miotkowski}, \citenamefont {Datta},\
  and\ \citenamefont {Chen}}]{tia17}%
  \BibitemOpen
  \bibfield  {author} {\bibinfo {author} {\bibfnamefont {J.}~\bibnamefont
  {Tian}}, \bibinfo {author} {\bibfnamefont {S.}~\bibnamefont {Hong}}, \bibinfo
  {author} {\bibfnamefont {I.}~\bibnamefont {Miotkowski}}, \bibinfo {author}
  {\bibfnamefont {S.}~\bibnamefont {Datta}},\ and\ \bibinfo {author}
  {\bibfnamefont {Y.~P.}\ \bibnamefont {Chen}},\ }\bibfield  {title} {\bibinfo
  {title} {Observation of current-induced, long-lived persistent spin
  polarization in a topological insulator: A rechargeable spin battery},\
  }\href {https://doi.org/10.1126/sciadv.1602531} {\bibfield  {journal}
  {\bibinfo  {journal} {Sci. Adv.}\ }\textbf {\bibinfo {volume} {3}},\ \bibinfo
  {pages} {e1602531} (\bibinfo {year} {2017})}\BibitemShut {NoStop}%
\bibitem [{\citenamefont {Xu}\ \emph {et~al.}(2017)\citenamefont {Xu},
  \citenamefont {Xu},\ and\ \citenamefont {Zhu}}]{xu17}%
  \BibitemOpen
  \bibfield  {author} {\bibinfo {author} {\bibfnamefont {N.}~\bibnamefont
  {Xu}}, \bibinfo {author} {\bibfnamefont {Y.}~\bibnamefont {Xu}},\ and\
  \bibinfo {author} {\bibfnamefont {J.}~\bibnamefont {Zhu}},\ }\bibfield
  {title} {\bibinfo {title} {Topological insulators for thermoelectrics},\
  }\href {https://doi.org/10.1038/s41535-017-0054-3} {\bibfield  {journal}
  {\bibinfo  {journal} {npj Quantum Mater.}\ }\textbf {\bibinfo {volume} {2}},\
  \bibinfo {pages} {51} (\bibinfo {year} {2017})}\BibitemShut {NoStop}%
\bibitem [{\citenamefont {Zhou}\ \emph {et~al.}(2008)\citenamefont {Zhou},
  \citenamefont {Lu}, \citenamefont {Chu}, \citenamefont {Shen},\ and\
  \citenamefont {Niu}}]{zho08}%
  \BibitemOpen
  \bibfield  {author} {\bibinfo {author} {\bibfnamefont {B.}~\bibnamefont
  {Zhou}}, \bibinfo {author} {\bibfnamefont {H.-Z.}\ \bibnamefont {Lu}},
  \bibinfo {author} {\bibfnamefont {R.-L.}\ \bibnamefont {Chu}}, \bibinfo
  {author} {\bibfnamefont {S.-Q.}\ \bibnamefont {Shen}},\ and\ \bibinfo
  {author} {\bibfnamefont {Q.}~\bibnamefont {Niu}},\ }\bibfield  {title}
  {\bibinfo {title} {Finite size effects on helical edge states in a quantum
  spin-{H}all system},\ }\href {https://doi.org/10.1103/PhysRevLett.101.246807}
  {\bibfield  {journal} {\bibinfo  {journal} {Phys. Rev. Lett.}\ }\textbf
  {\bibinfo {volume} {101}},\ \bibinfo {pages} {246807} (\bibinfo {year}
  {2008})}\BibitemShut {NoStop}%
\bibitem [{\citenamefont {Linder}\ \emph {et~al.}(2009)\citenamefont {Linder},
  \citenamefont {Yokoyama},\ and\ \citenamefont {Sudb\o{}}}]{lin09}%
  \BibitemOpen
  \bibfield  {author} {\bibinfo {author} {\bibfnamefont {J.}~\bibnamefont
  {Linder}}, \bibinfo {author} {\bibfnamefont {T.}~\bibnamefont {Yokoyama}},\
  and\ \bibinfo {author} {\bibfnamefont {A.}~\bibnamefont {Sudb\o{}}},\
  }\bibfield  {title} {\bibinfo {title} {Anomalous finite size effects on
  surface states in the topological insulator
  {${\text{Bi}}_{2}{\text{Se}}_{3}$}},\ }\href
  {https://doi.org/10.1103/PhysRevB.80.205401} {\bibfield  {journal} {\bibinfo
  {journal} {Phys. Rev. B}\ }\textbf {\bibinfo {volume} {80}},\ \bibinfo
  {pages} {205401} (\bibinfo {year} {2009})}\BibitemShut {NoStop}%
\bibitem [{\citenamefont {Liu}\ \emph {et~al.}(2010{\natexlab{a}})\citenamefont
  {Liu}, \citenamefont {Zhang}, \citenamefont {Yan}, \citenamefont {Qi},
  \citenamefont {Frauenheim}, \citenamefont {Dai}, \citenamefont {Fang},\ and\
  \citenamefont {Zhang}}]{liu10a}%
  \BibitemOpen
  \bibfield  {author} {\bibinfo {author} {\bibfnamefont {C.-X.}\ \bibnamefont
  {Liu}}, \bibinfo {author} {\bibfnamefont {H.}~\bibnamefont {Zhang}}, \bibinfo
  {author} {\bibfnamefont {B.}~\bibnamefont {Yan}}, \bibinfo {author}
  {\bibfnamefont {X.-L.}\ \bibnamefont {Qi}}, \bibinfo {author} {\bibfnamefont
  {T.}~\bibnamefont {Frauenheim}}, \bibinfo {author} {\bibfnamefont
  {X.}~\bibnamefont {Dai}}, \bibinfo {author} {\bibfnamefont {Z.}~\bibnamefont
  {Fang}},\ and\ \bibinfo {author} {\bibfnamefont {S.-C.}\ \bibnamefont
  {Zhang}},\ }\bibfield  {title} {\bibinfo {title} {Oscillatory crossover from
  two-dimensional to three-dimensional topological insulators},\ }\href
  {https://doi.org/10.1103/PhysRevB.81.041307} {\bibfield  {journal} {\bibinfo
  {journal} {Phys. Rev. B}\ }\textbf {\bibinfo {volume} {81}},\ \bibinfo
  {pages} {041307} (\bibinfo {year} {2010}{\natexlab{a}})}\BibitemShut
  {NoStop}%
\bibitem [{\citenamefont {Lu}\ \emph {et~al.}(2010)\citenamefont {Lu},
  \citenamefont {Shan}, \citenamefont {Yao}, \citenamefont {Niu},\ and\
  \citenamefont {Shen}}]{lu10}%
  \BibitemOpen
  \bibfield  {author} {\bibinfo {author} {\bibfnamefont {H.-Z.}\ \bibnamefont
  {Lu}}, \bibinfo {author} {\bibfnamefont {W.-Y.}\ \bibnamefont {Shan}},
  \bibinfo {author} {\bibfnamefont {W.}~\bibnamefont {Yao}}, \bibinfo {author}
  {\bibfnamefont {Q.}~\bibnamefont {Niu}},\ and\ \bibinfo {author}
  {\bibfnamefont {S.-Q.}\ \bibnamefont {Shen}},\ }\bibfield  {title} {\bibinfo
  {title} {Massive {D}irac fermions and spin physics in an ultrathin film of
  topological insulator},\ }\href {https://doi.org/10.1103/PhysRevB.81.115407}
  {\bibfield  {journal} {\bibinfo  {journal} {Phys. Rev. B}\ }\textbf {\bibinfo
  {volume} {81}},\ \bibinfo {pages} {115407} (\bibinfo {year}
  {2010})}\BibitemShut {NoStop}%
\bibitem [{\citenamefont {Kotulla}\ and\ \citenamefont
  {Z{\"u}licke}(2017)}]{kot17}%
  \BibitemOpen
  \bibfield  {author} {\bibinfo {author} {\bibfnamefont {M.}~\bibnamefont
  {Kotulla}}\ and\ \bibinfo {author} {\bibfnamefont {U.}~\bibnamefont
  {Z{\"u}licke}},\ }\bibfield  {title} {\bibinfo {title} {Manipulating
  topological-insulator properties using quantum confinement},\ }\href
  {https://doi.org/10.1088/1367-2630/aa7913} {\bibfield  {journal} {\bibinfo
  {journal} {New J. Phys.}\ }\textbf {\bibinfo {volume} {19}},\ \bibinfo
  {pages} {073025} (\bibinfo {year} {2017})}\BibitemShut {NoStop}%
\bibitem [{\citenamefont {Gioia}\ \emph {et~al.}(2018)\citenamefont {Gioia},
  \citenamefont {Z\"ulicke}, \citenamefont {Governale},\ and\ \citenamefont
  {Winkler}}]{gio18}%
  \BibitemOpen
  \bibfield  {author} {\bibinfo {author} {\bibfnamefont {L.}~\bibnamefont
  {Gioia}}, \bibinfo {author} {\bibfnamefont {U.}~\bibnamefont {Z\"ulicke}},
  \bibinfo {author} {\bibfnamefont {M.}~\bibnamefont {Governale}},\ and\
  \bibinfo {author} {\bibfnamefont {R.}~\bibnamefont {Winkler}},\ }\bibfield
  {title} {\bibinfo {title} {Dirac electrons in quantum rings},\ }\href
  {https://doi.org/10.1103/PhysRevB.97.205421} {\bibfield  {journal} {\bibinfo
  {journal} {Phys. Rev. B}\ }\textbf {\bibinfo {volume} {97}},\ \bibinfo
  {pages} {205421} (\bibinfo {year} {2018})}\BibitemShut {NoStop}%
\bibitem [{\citenamefont {Malkova}\ and\ \citenamefont {Bryant}(2010)}]{mal10}%
  \BibitemOpen
  \bibfield  {author} {\bibinfo {author} {\bibfnamefont {N.}~\bibnamefont
  {Malkova}}\ and\ \bibinfo {author} {\bibfnamefont {G.~W.}\ \bibnamefont
  {Bryant}},\ }\bibfield  {title} {\bibinfo {title} {Negative-band-gap quantum
  dots: {G}ap collapse, intrinsic surface states, excitonic response, and
  excitonic insulator phase},\ }\href
  {https://doi.org/10.1103/PhysRevB.82.155314} {\bibfield  {journal} {\bibinfo
  {journal} {Phys. Rev. B}\ }\textbf {\bibinfo {volume} {82}},\ \bibinfo
  {pages} {155314} (\bibinfo {year} {2010})}\BibitemShut {NoStop}%
\bibitem [{\citenamefont {Imura}\ \emph {et~al.}(2012)\citenamefont {Imura},
  \citenamefont {Yoshimura}, \citenamefont {Takane},\ and\ \citenamefont
  {Fukui}}]{imu12}%
  \BibitemOpen
  \bibfield  {author} {\bibinfo {author} {\bibfnamefont {K.-I.}\ \bibnamefont
  {Imura}}, \bibinfo {author} {\bibfnamefont {Y.}~\bibnamefont {Yoshimura}},
  \bibinfo {author} {\bibfnamefont {Y.}~\bibnamefont {Takane}},\ and\ \bibinfo
  {author} {\bibfnamefont {T.}~\bibnamefont {Fukui}},\ }\bibfield  {title}
  {\bibinfo {title} {Spherical topological insulator},\ }\href
  {https://doi.org/10.1103/PhysRevB.86.235119} {\bibfield  {journal} {\bibinfo
  {journal} {Phys. Rev. B}\ }\textbf {\bibinfo {volume} {86}},\ \bibinfo
  {pages} {235119} (\bibinfo {year} {2012})}\BibitemShut {NoStop}%
\bibitem [{\citenamefont {Takane}\ and\ \citenamefont {Imura}(2013)}]{tak13}%
  \BibitemOpen
  \bibfield  {author} {\bibinfo {author} {\bibfnamefont {Y.}~\bibnamefont
  {Takane}}\ and\ \bibinfo {author} {\bibfnamefont {K.-I.}\ \bibnamefont
  {Imura}},\ }\bibfield  {title} {\bibinfo {title} {Unified description of
  {D}irac electrons on a curved surface of topological insulators},\ }\href
  {https://doi.org/10.7566/JPSJ.82.074712} {\bibfield  {journal} {\bibinfo
  {journal} {J. Phys. Soc. Jpn.}\ }\textbf {\bibinfo {volume} {82}},\ \bibinfo
  {pages} {074712} (\bibinfo {year} {2013})}\BibitemShut {NoStop}%
\bibitem [{\citenamefont {Siroki}\ \emph {et~al.}(2016)\citenamefont {Siroki},
  \citenamefont {Lee}, \citenamefont {Haynes},\ and\ \citenamefont
  {Giannini}}]{sir16}%
  \BibitemOpen
  \bibfield  {author} {\bibinfo {author} {\bibfnamefont {G.}~\bibnamefont
  {Siroki}}, \bibinfo {author} {\bibfnamefont {D.~K.~K.}\ \bibnamefont {Lee}},
  \bibinfo {author} {\bibfnamefont {P.~D.}\ \bibnamefont {Haynes}},\ and\
  \bibinfo {author} {\bibfnamefont {V.}~\bibnamefont {Giannini}},\ }\bibfield
  {title} {\bibinfo {title} {Single-electron induced surface plasmons on a
  topological nanoparticle},\ }\href {https://doi.org/10.1038/ncomms12375}
  {\bibfield  {journal} {\bibinfo  {journal} {Nat. Commun.}\ }\textbf {\bibinfo
  {volume} {7}},\ \bibinfo {pages} {12375} (\bibinfo {year}
  {2016})}\BibitemShut {NoStop}%
\bibitem [{\citenamefont {Siroki}\ \emph {et~al.}(2017)\citenamefont {Siroki},
  \citenamefont {Haynes}, \citenamefont {Lee},\ and\ \citenamefont
  {Giannini}}]{sir17}%
  \BibitemOpen
  \bibfield  {author} {\bibinfo {author} {\bibfnamefont {G.}~\bibnamefont
  {Siroki}}, \bibinfo {author} {\bibfnamefont {P.~D.}\ \bibnamefont {Haynes}},
  \bibinfo {author} {\bibfnamefont {D.~K.~K.}\ \bibnamefont {Lee}},\ and\
  \bibinfo {author} {\bibfnamefont {V.}~\bibnamefont {Giannini}},\ }\bibfield
  {title} {\bibinfo {title} {Protection of surface states in topological
  nanoparticles},\ }\href {https://doi.org/10.1103/PhysRevMaterials.1.024201}
  {\bibfield  {journal} {\bibinfo  {journal} {Phys. Rev. Materials}\ }\textbf
  {\bibinfo {volume} {1}},\ \bibinfo {pages} {024201} (\bibinfo {year}
  {2017})}\BibitemShut {NoStop}%
\bibitem [{\citenamefont {Zirnstein}\ and\ \citenamefont
  {Rosenow}(2017)}]{zir17}%
  \BibitemOpen
  \bibfield  {author} {\bibinfo {author} {\bibfnamefont {H.-G.}\ \bibnamefont
  {Zirnstein}}\ and\ \bibinfo {author} {\bibfnamefont {B.}~\bibnamefont
  {Rosenow}},\ }\bibfield  {title} {\bibinfo {title} {Time-reversal-symmetric
  topological magnetoelectric effect in three-dimensional topological
  insulators},\ }\href {https://doi.org/10.1103/PhysRevB.96.201112} {\bibfield
  {journal} {\bibinfo  {journal} {Phys. Rev. B}\ }\textbf {\bibinfo {volume}
  {96}},\ \bibinfo {pages} {201112} (\bibinfo {year} {2017})}\BibitemShut
  {NoStop}%
\bibitem [{\citenamefont {Neupert}\ \emph {et~al.}(2015)\citenamefont
  {Neupert}, \citenamefont {Rachel}, \citenamefont {Thomale},\ and\
  \citenamefont {Greiter}}]{neu15}%
  \BibitemOpen
  \bibfield  {author} {\bibinfo {author} {\bibfnamefont {T.}~\bibnamefont
  {Neupert}}, \bibinfo {author} {\bibfnamefont {S.}~\bibnamefont {Rachel}},
  \bibinfo {author} {\bibfnamefont {R.}~\bibnamefont {Thomale}},\ and\ \bibinfo
  {author} {\bibfnamefont {M.}~\bibnamefont {Greiter}},\ }\bibfield  {title}
  {\bibinfo {title} {Interacting surface states of three-dimensional
  topological insulators},\ }\href
  {https://doi.org/10.1103/PhysRevLett.115.017001} {\bibfield  {journal}
  {\bibinfo  {journal} {Phys. Rev. Lett.}\ }\textbf {\bibinfo {volume} {115}},\
  \bibinfo {pages} {017001} (\bibinfo {year} {2015})}\BibitemShut {NoStop}%
\bibitem [{\citenamefont {Durst}(2016)}]{dur16}%
  \BibitemOpen
  \bibfield  {author} {\bibinfo {author} {\bibfnamefont {A.~C.}\ \bibnamefont
  {Durst}},\ }\bibfield  {title} {\bibinfo {title} {Disorder-induced density of
  states on the surface of a spherical topological insulator},\ }\href
  {https://doi.org/10.1103/PhysRevB.93.245424} {\bibfield  {journal} {\bibinfo
  {journal} {Phys. Rev. B}\ }\textbf {\bibinfo {volume} {93}},\ \bibinfo
  {pages} {245424} (\bibinfo {year} {2016})}\BibitemShut {NoStop}%
\bibitem [{\citenamefont {Abrikosov}(2002)}]{abr02}%
  \BibitemOpen
  \bibfield  {author} {\bibinfo {author} {\bibfnamefont {A.~A.}\ \bibnamefont
  {Abrikosov}},\ }\bibfield  {title} {\bibinfo {title} {Fermion states on the
  sphere {$S^2$}},\ }\href {https://doi.org/10.1142/S0217751X02010261}
  {\bibfield  {journal} {\bibinfo  {journal} {Int. J. Mod. Phys. A}\ }\textbf
  {\bibinfo {volume} {17}},\ \bibinfo {pages} {885} (\bibinfo {year} {2002})},\
  \bibinfo {note} {see arXiv:hep-th/0212134 for an extended version that also
  contains a useful collection of relevant references.}\BibitemShut {Stop}%
\bibitem [{\citenamefont {Lee}(2009)}]{lee09}%
  \BibitemOpen
  \bibfield  {author} {\bibinfo {author} {\bibfnamefont {D.-H.}\ \bibnamefont
  {Lee}},\ }\bibfield  {title} {\bibinfo {title} {Surface states of topological
  insulators: The {D}irac fermion in curved two-dimensional spaces},\ }\href
  {https://doi.org/10.1103/PhysRevLett.103.196804} {\bibfield  {journal}
  {\bibinfo  {journal} {Phys. Rev. Lett.}\ }\textbf {\bibinfo {volume} {103}},\
  \bibinfo {pages} {196804} (\bibinfo {year} {2009})}\BibitemShut {NoStop}%
\bibitem [{\citenamefont {Parente}\ \emph {et~al.}(2011)\citenamefont
  {Parente}, \citenamefont {Lucignano}, \citenamefont {Vitale}, \citenamefont
  {Tagliacozzo},\ and\ \citenamefont {Guinea}}]{par11}%
  \BibitemOpen
  \bibfield  {author} {\bibinfo {author} {\bibfnamefont {V.}~\bibnamefont
  {Parente}}, \bibinfo {author} {\bibfnamefont {P.}~\bibnamefont {Lucignano}},
  \bibinfo {author} {\bibfnamefont {P.}~\bibnamefont {Vitale}}, \bibinfo
  {author} {\bibfnamefont {A.}~\bibnamefont {Tagliacozzo}},\ and\ \bibinfo
  {author} {\bibfnamefont {F.}~\bibnamefont {Guinea}},\ }\bibfield  {title}
  {\bibinfo {title} {Spin connection and boundary states in a topological
  insulator},\ }\href {https://doi.org/10.1103/PhysRevB.83.075424} {\bibfield
  {journal} {\bibinfo  {journal} {Phys. Rev. B}\ }\textbf {\bibinfo {volume}
  {83}},\ \bibinfo {pages} {075424} (\bibinfo {year} {2011})}\BibitemShut
  {NoStop}%
\bibitem [{\citenamefont {Layeghnejad}\ \emph {et~al.}(2011)\citenamefont
  {Layeghnejad}, \citenamefont {Zare},\ and\ \citenamefont {Moazzemi}}]{lay11}%
  \BibitemOpen
  \bibfield  {author} {\bibinfo {author} {\bibfnamefont {R.}~\bibnamefont
  {Layeghnejad}}, \bibinfo {author} {\bibfnamefont {M.}~\bibnamefont {Zare}},\
  and\ \bibinfo {author} {\bibfnamefont {R.}~\bibnamefont {Moazzemi}},\
  }\bibfield  {title} {\bibinfo {title} {Dirac particle in a spherical scalar
  potential well},\ }\href {https://doi.org/10.1103/PhysRevD.84.125026}
  {\bibfield  {journal} {\bibinfo  {journal} {Phys. Rev. D}\ }\textbf {\bibinfo
  {volume} {84}},\ \bibinfo {pages} {125026} (\bibinfo {year}
  {2011})}\BibitemShut {NoStop}%
\bibitem [{\citenamefont {Paudel}\ and\ \citenamefont
  {Leuenberger}(2013)}]{pau13}%
  \BibitemOpen
  \bibfield  {author} {\bibinfo {author} {\bibfnamefont {H.~P.}\ \bibnamefont
  {Paudel}}\ and\ \bibinfo {author} {\bibfnamefont {M.~N.}\ \bibnamefont
  {Leuenberger}},\ }\bibfield  {title} {\bibinfo {title} {Three-dimensional
  topological insulator quantum dot for optically controlled quantum memory and
  quantum computing},\ }\href {https://doi.org/10.1103/PhysRevB.88.085316}
  {\bibfield  {journal} {\bibinfo  {journal} {Phys. Rev. B}\ }\textbf {\bibinfo
  {volume} {88}},\ \bibinfo {pages} {085316} (\bibinfo {year}
  {2013})}\BibitemShut {NoStop}%
\bibitem [{\citenamefont {Zhang}\ \emph {et~al.}(2009)\citenamefont {Zhang},
  \citenamefont {Liu}, \citenamefont {Qi}, \citenamefont {Dai}, \citenamefont
  {Fang},\ and\ \citenamefont {Zhang}}]{zha09}%
  \BibitemOpen
  \bibfield  {author} {\bibinfo {author} {\bibfnamefont {H.}~\bibnamefont
  {Zhang}}, \bibinfo {author} {\bibfnamefont {C.-X.}\ \bibnamefont {Liu}},
  \bibinfo {author} {\bibfnamefont {X.-L.}\ \bibnamefont {Qi}}, \bibinfo
  {author} {\bibfnamefont {X.}~\bibnamefont {Dai}}, \bibinfo {author}
  {\bibfnamefont {Z.}~\bibnamefont {Fang}},\ and\ \bibinfo {author}
  {\bibfnamefont {S.-C.}\ \bibnamefont {Zhang}},\ }\bibfield  {title} {\bibinfo
  {title} {Topological insulators in {$\mathrm{Bi}_2\mathrm{Se}_3$},
  {$\mathrm{Bi}_2\mathrm{Te}_3$}, and {$\mathrm{Sb}_2\mathrm{Te}_3$} with a
  single {D}irac cone on the surface},\ }\href
  {https://doi.org/10.1038/nphys1270} {\bibfield  {journal} {\bibinfo
  {journal} {Nat. Phys.}\ }\textbf {\bibinfo {volume} {82}},\ \bibinfo {pages}
  {438} (\bibinfo {year} {2009})}\BibitemShut {NoStop}%
\bibitem [{\citenamefont {Liu}\ \emph {et~al.}(2010{\natexlab{b}})\citenamefont
  {Liu}, \citenamefont {Qi}, \citenamefont {Zhang}, \citenamefont {Dai},
  \citenamefont {Fang},\ and\ \citenamefont {Zhang}}]{liu10}%
  \BibitemOpen
  \bibfield  {author} {\bibinfo {author} {\bibfnamefont {C.-X.}\ \bibnamefont
  {Liu}}, \bibinfo {author} {\bibfnamefont {X.-L.}\ \bibnamefont {Qi}},
  \bibinfo {author} {\bibfnamefont {H.}~\bibnamefont {Zhang}}, \bibinfo
  {author} {\bibfnamefont {X.}~\bibnamefont {Dai}}, \bibinfo {author}
  {\bibfnamefont {Z.}~\bibnamefont {Fang}},\ and\ \bibinfo {author}
  {\bibfnamefont {S.-C.}\ \bibnamefont {Zhang}},\ }\bibfield  {title} {\bibinfo
  {title} {Model hamiltonian for topological insulators},\ }\href
  {https://doi.org/10.1103/PhysRevB.82.045122} {\bibfield  {journal} {\bibinfo
  {journal} {Phys. Rev. B}\ }\textbf {\bibinfo {volume} {82}},\ \bibinfo
  {pages} {045122} (\bibinfo {year} {2010}{\natexlab{b}})}\BibitemShut
  {NoStop}%
\bibitem [{\citenamefont {Brems}\ \emph {et~al.}(2018)\citenamefont {Brems},
  \citenamefont {Paaske}, \citenamefont {Lunde},\ and\ \citenamefont
  {Willatzen}}]{bre18}%
  \BibitemOpen
  \bibfield  {author} {\bibinfo {author} {\bibfnamefont {M.~R.}\ \bibnamefont
  {Brems}}, \bibinfo {author} {\bibfnamefont {J.}~\bibnamefont {Paaske}},
  \bibinfo {author} {\bibfnamefont {A.~M.}\ \bibnamefont {Lunde}},\ and\
  \bibinfo {author} {\bibfnamefont {M.}~\bibnamefont {Willatzen}},\ }\bibfield
  {title} {\bibinfo {title} {Symmetry analysis of strain, electric and magnetic
  fields in the {Bi$_2$Se$_3$}-class of topological insulators},\ }\href
  {https://doi.org/10.1088/1367-2630/aabcfc} {\bibfield  {journal} {\bibinfo
  {journal} {New J. Phys.}\ }\textbf {\bibinfo {volume} {20}},\ \bibinfo
  {pages} {053041} (\bibinfo {year} {2018})}\BibitemShut {NoStop}%
\bibitem [{not()}]{notePauli}%
  \BibitemOpen
  \href@noop {} {}\bibinfo {note} {The $\tau_j$ ($\sigma_j$) for $j=x, y, z$
  are standard Pauli matrices for the pseudo-spin-$1/2$ (real-spin-$1/2$)
  degree of freedom, and $\tau_0$ ($\sigma_0$) is the identity operator in that
  subspace. The symbol $\otimes$ indicates the direct product of operators from
  pseudo-spin and real-spin spaces. The combinations $\sigma_\pm\equiv
  (\sigma_x\pm i\sigma_y)/2$ are ladder operators for real spin, and
  $\vek{\sigma}$ denotes the vector $(\sigma_x, \sigma_y,
  \sigma_z)$.}\BibitemShut {Stop}%
\bibitem [{\citenamefont {Nechaev}\ and\ \citenamefont
  {Krasovskii}(2016)}]{nec16}%
  \BibitemOpen
  \bibfield  {author} {\bibinfo {author} {\bibfnamefont {I.~A.}\ \bibnamefont
  {Nechaev}}\ and\ \bibinfo {author} {\bibfnamefont {E.~E.}\ \bibnamefont
  {Krasovskii}},\ }\bibfield  {title} {\bibinfo {title} {Relativistic
  $\mathrm{k}\cdot\mathrm{p}$ {H}amiltonians for centrosymmetric topological
  insulators from \textit{ab initio} wave functions},\ }\href
  {https://doi.org/10.1103/PhysRevB.94.201410} {\bibfield  {journal} {\bibinfo
  {journal} {Phys. Rev. B}\ }\textbf {\bibinfo {volume} {94}},\ \bibinfo
  {pages} {201410} (\bibinfo {year} {2016})}\BibitemShut {NoStop}%
\bibitem [{\citenamefont {Novik}\ \emph {et~al.}(2005)\citenamefont {Novik},
  \citenamefont {Pfeuffer-Jeschke}, \citenamefont {Jungwirth}, \citenamefont
  {Latussek}, \citenamefont {Becker}, \citenamefont {Landwehr}, \citenamefont
  {Buhmann},\ and\ \citenamefont {Molenkamp}}]{nov05}%
  \BibitemOpen
  \bibfield  {author} {\bibinfo {author} {\bibfnamefont {E.~G.}\ \bibnamefont
  {Novik}}, \bibinfo {author} {\bibfnamefont {A.}~\bibnamefont
  {Pfeuffer-Jeschke}}, \bibinfo {author} {\bibfnamefont {T.}~\bibnamefont
  {Jungwirth}}, \bibinfo {author} {\bibfnamefont {V.}~\bibnamefont {Latussek}},
  \bibinfo {author} {\bibfnamefont {C.~R.}\ \bibnamefont {Becker}}, \bibinfo
  {author} {\bibfnamefont {G.}~\bibnamefont {Landwehr}}, \bibinfo {author}
  {\bibfnamefont {H.}~\bibnamefont {Buhmann}},\ and\ \bibinfo {author}
  {\bibfnamefont {L.~W.}\ \bibnamefont {Molenkamp}},\ }\bibfield  {title}
  {\bibinfo {title} {Band structure of semimagnetic
  {${\mathrm{Hg}}_{1\ensuremath{-}y}{\mathrm{Mn}}_{y}\mathrm{Te}$} quantum
  wells},\ }\href {https://doi.org/10.1103/PhysRevB.72.035321} {\bibfield
  {journal} {\bibinfo  {journal} {Phys. Rev. B}\ }\textbf {\bibinfo {volume}
  {72}},\ \bibinfo {pages} {035321} (\bibinfo {year} {2005})}\BibitemShut
  {NoStop}%
\bibitem [{\citenamefont {Klipstein}(2015)}]{kli15}%
  \BibitemOpen
  \bibfield  {author} {\bibinfo {author} {\bibfnamefont {P.~C.}\ \bibnamefont
  {Klipstein}},\ }\bibfield  {title} {\bibinfo {title} {Structure of the
  quantum spin {H}all states in {HgTe/CdTe and InAs/GaSb/AlSb} quantum wells},\
  }\href {https://doi.org/10.1103/PhysRevB.91.035310} {\bibfield  {journal}
  {\bibinfo  {journal} {Phys. Rev. B}\ }\textbf {\bibinfo {volume} {91}},\
  \bibinfo {pages} {035310} (\bibinfo {year} {2015})}\BibitemShut {NoStop}%
\bibitem [{\citenamefont {Assaf}\ \emph {et~al.}(2017)\citenamefont {Assaf},
  \citenamefont {Phuphachong}, \citenamefont {Volobuev}, \citenamefont {Bauer},
  \citenamefont {Springholz}, \citenamefont {de~Vaulchier},\ and\ \citenamefont
  {Guldner}}]{ass17}%
  \BibitemOpen
  \bibfield  {author} {\bibinfo {author} {\bibfnamefont {B.~A.}\ \bibnamefont
  {Assaf}}, \bibinfo {author} {\bibfnamefont {T.}~\bibnamefont {Phuphachong}},
  \bibinfo {author} {\bibfnamefont {V.~V.}\ \bibnamefont {Volobuev}}, \bibinfo
  {author} {\bibfnamefont {G.}~\bibnamefont {Bauer}}, \bibinfo {author}
  {\bibfnamefont {G.}~\bibnamefont {Springholz}}, \bibinfo {author}
  {\bibfnamefont {L.-A.}\ \bibnamefont {de~Vaulchier}},\ and\ \bibinfo {author}
  {\bibfnamefont {Y.}~\bibnamefont {Guldner}},\ }\bibfield  {title} {\bibinfo
  {title} {Magnetooptical determination of a topological index},\ }\href
  {https://doi.org/10.1038/s41535-017-0028-5} {\bibfield  {journal} {\bibinfo
  {journal} {npj Quant. Mater.}\ }\textbf {\bibinfo {volume} {2}},\ \bibinfo
  {pages} {26} (\bibinfo {year} {2017})}\BibitemShut {NoStop}%
\bibitem [{\citenamefont {Assaf}\ \emph {et~al.}(2016)\citenamefont {Assaf},
  \citenamefont {Phuphachong}, \citenamefont {Volobuev}, \citenamefont
  {Inhofer}, \citenamefont {Bauer}, \citenamefont {Springholz}, \citenamefont
  {de~Vaulchier},\ and\ \citenamefont {Guldner}}]{ass16}%
  \BibitemOpen
  \bibfield  {author} {\bibinfo {author} {\bibfnamefont {B.~A.}\ \bibnamefont
  {Assaf}}, \bibinfo {author} {\bibfnamefont {T.}~\bibnamefont {Phuphachong}},
  \bibinfo {author} {\bibfnamefont {V.~V.}\ \bibnamefont {Volobuev}}, \bibinfo
  {author} {\bibfnamefont {A.}~\bibnamefont {Inhofer}}, \bibinfo {author}
  {\bibfnamefont {G.}~\bibnamefont {Bauer}}, \bibinfo {author} {\bibfnamefont
  {G.}~\bibnamefont {Springholz}}, \bibinfo {author} {\bibfnamefont {L.~A.}\
  \bibnamefont {de~Vaulchier}},\ and\ \bibinfo {author} {\bibfnamefont
  {Y.}~\bibnamefont {Guldner}},\ }\bibfield  {title} {\bibinfo {title} {Massive
  and massless {D}irac fermions in {Pb$_{1-x}$Sn$_x$ Te} topological
  crystalline insulator probed by magneto-optical absorption},\ }\href
  {https://doi.org/10.1038/srep20323} {\bibfield  {journal} {\bibinfo
  {journal} {Sci. Rep.}\ }\textbf {\bibinfo {volume} {6}},\ \bibinfo {pages}
  {20323} (\bibinfo {year} {2016})}\BibitemShut {NoStop}%
\bibitem [{\citenamefont {Sakurai}\ and\ \citenamefont
  {Napolitano}(2011)}]{sak11}%
  \BibitemOpen
  \bibfield  {author} {\bibinfo {author} {\bibfnamefont {J.~J.}\ \bibnamefont
  {Sakurai}}\ and\ \bibinfo {author} {\bibfnamefont {J.}~\bibnamefont
  {Napolitano}},\ }\href@noop {} {\emph {\bibinfo {title} {Modern Quantum
  Mechanics}}},\ \bibinfo {edition} {2nd}\ ed.\ (\bibinfo  {publisher}
  {Addison-Wesley},\ \bibinfo {address} {San Francisco},\ \bibinfo {year}
  {2011})\BibitemShut {NoStop}%
\bibitem [{\citenamefont {Sheka}\ and\ \citenamefont {Sheka}(1966)}]{she67}%
  \BibitemOpen
  \bibfield  {author} {\bibinfo {author} {\bibfnamefont {V.~I.}\ \bibnamefont
  {Sheka}}\ and\ \bibinfo {author} {\bibfnamefont {D.~I.}\ \bibnamefont
  {Sheka}},\ }\bibfield  {title} {\bibinfo {title} {Local states in a
  semiconductor with a narrow forbidden band},\ }\href@noop {} {\bibfield
  {journal} {\bibinfo  {journal} {Zh. Eksp. Teor. Fiz.}\ }\textbf {\bibinfo
  {volume} {51}},\ \bibinfo {pages} {1445} (\bibinfo {year} {1966})},\ \bibinfo
  {note} {[Sov. Phys. JETP \textbf{24}, 975 (1967)]}\BibitemShut {NoStop}%
\bibitem [{\citenamefont {Efros}\ and\ \citenamefont {Rosen}(1998)}]{efr98}%
  \BibitemOpen
  \bibfield  {author} {\bibinfo {author} {\bibfnamefont {A.~L.}\ \bibnamefont
  {Efros}}\ and\ \bibinfo {author} {\bibfnamefont {M.}~\bibnamefont {Rosen}},\
  }\bibfield  {title} {\bibinfo {title} {Quantum size level structure of
  narrow-gap semiconductor nanocrystals: {E}ffect of band coupling},\ }\href
  {https://doi.org/10.1103/PhysRevB.58.7120} {\bibfield  {journal} {\bibinfo
  {journal} {Phys. Rev. B}\ }\textbf {\bibinfo {volume} {58}},\ \bibinfo
  {pages} {7120} (\bibinfo {year} {1998})}\BibitemShut {NoStop}%
\bibitem [{\citenamefont {Thaller}(1992)}]{tha92}%
  \BibitemOpen
  \bibfield  {author} {\bibinfo {author} {\bibfnamefont {B.}~\bibnamefont
  {Thaller}},\ }\href {https://doi.org/10.1007/978-3-662-02753-0} {\emph
  {\bibinfo {title} {The Dirac Equation}}}\ (\bibinfo  {publisher} {Springer},\
  \bibinfo {address} {Berlin},\ \bibinfo {year} {1992})\BibitemShut {NoStop}%
\bibitem [{Note1()}]{Note1}%
  \BibitemOpen
  \bibinfo {note} {Within a conventional nomenclature~\cite {hau04}, the term
  proportional to the position vector ${\protect \ensuremath {\protect \bm
  {\protect \mathrm {r}}}}$ (the TI-material Compton length $R_0$) in Eq.~(\ref
  {eq:envDip}) is associated with intra-band (inter-band) transitions. This
  jargon is somewhat misleading in the case of narrow-gap (e.g., TI)
  materials~\cite {pau13}. More precisely, the first (second) term on the
  r.h.s.\ of Eq.~(\ref {eq:envDip}) pertains to dipole transitions mediated by
  the envelope part (the Bloch-function basis) of the confined-electron
  states.}\BibitemShut {Stop}%
\bibitem [{\citenamefont {Greiner}(1998)}]{gre98}%
  \BibitemOpen
  \bibfield  {author} {\bibinfo {author} {\bibfnamefont {W.}~\bibnamefont
  {Greiner}},\ }\href@noop {} {\emph {\bibinfo {title} {Quantum Mechanics
  Special Chapters}}}\ (\bibinfo  {publisher} {Springer},\ \bibinfo {address}
  {Berlin},\ \bibinfo {year} {1998})\BibitemShut {NoStop}%
\bibitem [{Note2()}]{Note2}%
  \BibitemOpen
  \bibinfo {note} {The degree of freedom associated with the quantum number
  $\sigma $ is the projection of a real spin-$1/2$ angular momentum, but this
  is not the actual electron spin~\cite {bre18}.}\BibitemShut {Stop}%
\bibitem [{\citenamefont {Varshalovich}\ \emph {et~al.}(1988)\citenamefont
  {Varshalovich}, \citenamefont {Moskalev},\ and\ \citenamefont
  {Khersonskii}}]{var88}%
  \BibitemOpen
  \bibfield  {author} {\bibinfo {author} {\bibfnamefont {D.~A.}\ \bibnamefont
  {Varshalovich}}, \bibinfo {author} {\bibfnamefont {A.~N.}\ \bibnamefont
  {Moskalev}},\ and\ \bibinfo {author} {\bibfnamefont {V.~K.}\ \bibnamefont
  {Khersonskii}},\ }\href {https://doi.org/10.1142/0270} {\emph {\bibinfo
  {title} {Quantum Theory of Angular Momentum}}}\ (\bibinfo  {publisher} {World
  Scientific},\ \bibinfo {address} {Singapore},\ \bibinfo {year} {1988})\
  \bibinfo {note} {{S}ection 7.2}\BibitemShut {NoStop}%
\bibitem [{\citenamefont {Szmytkowski}(2007)}]{szm07}%
  \BibitemOpen
  \bibfield  {author} {\bibinfo {author} {\bibfnamefont {R.}~\bibnamefont
  {Szmytkowski}},\ }\bibfield  {title} {\bibinfo {title} {Recurrence and
  differential relations for spherical spinors},\ }\href
  {https://doi.org/10.1007/s10910-006-9110-0} {\bibfield  {journal} {\bibinfo
  {journal} {J. Math. Chem.}\ }\textbf {\bibinfo {volume} {42}},\ \bibinfo
  {pages} {397} (\bibinfo {year} {2007})}\BibitemShut {NoStop}%
\bibitem [{\citenamefont {Abramowitz}\ and\ \citenamefont
  {Stegun}(1964)}]{abr64}%
  \BibitemOpen
  \bibfield  {author} {\bibinfo {author} {\bibfnamefont {M.}~\bibnamefont
  {Abramowitz}}\ and\ \bibinfo {author} {\bibfnamefont {I.~A.}\ \bibnamefont
  {Stegun}},\ }\href@noop {} {\emph {\bibinfo {title} {Handbook of Mathematical
  Functions}}}\ (\bibinfo  {publisher} {Dover},\ \bibinfo {address} {New
  York},\ \bibinfo {year} {1964})\BibitemShut {NoStop}%
\bibitem [{\citenamefont {Alberto}\ \emph {et~al.}(1996)\citenamefont
  {Alberto}, \citenamefont {Fiolhais},\ and\ \citenamefont {Gil}}]{alb96}%
  \BibitemOpen
  \bibfield  {author} {\bibinfo {author} {\bibfnamefont {P.}~\bibnamefont
  {Alberto}}, \bibinfo {author} {\bibfnamefont {C.}~\bibnamefont {Fiolhais}},\
  and\ \bibinfo {author} {\bibfnamefont {V.~M.~S.}\ \bibnamefont {Gil}},\
  }\bibfield  {title} {\bibinfo {title} {Relativistic particle in a box},\
  }\href {https://doi.org/10.1088/0143-0807/17/1/004} {\bibfield  {journal}
  {\bibinfo  {journal} {Eur. J. Phys.}\ }\textbf {\bibinfo {volume} {17}},\
  \bibinfo {pages} {19} (\bibinfo {year} {1996})}\BibitemShut {NoStop}%
\bibitem [{\citenamefont {Bernevig}\ \emph {et~al.}(2006)\citenamefont
  {Bernevig}, \citenamefont {Hughes},\ and\ \citenamefont {Zhang}}]{ber06}%
  \BibitemOpen
  \bibfield  {author} {\bibinfo {author} {\bibfnamefont {B.~A.}\ \bibnamefont
  {Bernevig}}, \bibinfo {author} {\bibfnamefont {T.~L.}\ \bibnamefont
  {Hughes}},\ and\ \bibinfo {author} {\bibfnamefont {S.}~\bibnamefont
  {Zhang}},\ }\bibfield  {title} {\bibinfo {title} {Quantum spin {H}all effect
  and topological phase transition in {HgTe} quantum wells},\ }\href
  {https://doi.org/10.1126/science.1133734} {\bibfield  {journal} {\bibinfo
  {journal} {Science}\ }\textbf {\bibinfo {volume} {314}},\ \bibinfo {pages}
  {1757} (\bibinfo {year} {2006})}\BibitemShut {NoStop}%
\bibitem [{\citenamefont {White}\ and\ \citenamefont {Sham}(1981)}]{whi81}%
  \BibitemOpen
  \bibfield  {author} {\bibinfo {author} {\bibfnamefont {S.~R.}\ \bibnamefont
  {White}}\ and\ \bibinfo {author} {\bibfnamefont {L.~J.}\ \bibnamefont
  {Sham}},\ }\bibfield  {title} {\bibinfo {title} {Electronic properties of
  flat-band semiconductor heterostructures},\ }\href
  {https://doi.org/10.1103/PhysRevLett.47.879} {\bibfield  {journal} {\bibinfo
  {journal} {Phys. Rev. Lett.}\ }\textbf {\bibinfo {volume} {47}},\ \bibinfo
  {pages} {879} (\bibinfo {year} {1981})}\BibitemShut {NoStop}%
\bibitem [{\citenamefont {Schuurmans}\ and\ \citenamefont
  {'t~Hooft}(1985)}]{sch85}%
  \BibitemOpen
  \bibfield  {author} {\bibinfo {author} {\bibfnamefont {M.~F.~H.}\
  \bibnamefont {Schuurmans}}\ and\ \bibinfo {author} {\bibfnamefont {G.~W.}\
  \bibnamefont {'t~Hooft}},\ }\bibfield  {title} {\bibinfo {title} {Simple
  calculations of confinement states in a quantum well},\ }\href
  {https://doi.org/10.1103/PhysRevB.31.8041} {\bibfield  {journal} {\bibinfo
  {journal} {Phys. Rev. B}\ }\textbf {\bibinfo {volume} {31}},\ \bibinfo
  {pages} {8041} (\bibinfo {year} {1985})}\BibitemShut {NoStop}%
\bibitem [{\citenamefont {Klipstein}(2018)}]{kli18}%
  \BibitemOpen
  \bibfield  {author} {\bibinfo {author} {\bibfnamefont {P.~C.}\ \bibnamefont
  {Klipstein}},\ }\bibfield  {title} {\bibinfo {title} {A $\mathbf{k\cdot p}$
  treatment of edge states in narrow {2D} topological insulators, with standard
  boundary conditions for the wave function and its derivative},\ }\href
  {https://doi.org/10.1088/1361-648X/aac85b} {\bibfield  {journal} {\bibinfo
  {journal} {J. Phys.: Condens. Matter}\ }\textbf {\bibinfo {volume} {30}},\
  \bibinfo {pages} {275302} (\bibinfo {year} {2018})}\BibitemShut {NoStop}%
\bibitem [{Note3()}]{Note3}%
  \BibitemOpen
  \bibinfo {note} {We find the expressions for $\protect \mathaccentV
  {tilde}07Ek$ and $\protect \mathaccentV {tilde}07Eq$ by solving for these in
  the relations ${\protect \mathaccentV {tilde}07EE}^{(\protect \mathrm
  {D})}_\pm (k) = \protect \mathaccentV {tilde}07EE$ and ${\protect
  \mathaccentV {tilde}07EE}^{(\protect \mathrm {B})}_\pm (q) = \protect
  \mathaccentV {tilde}07EE$, respectively. Then $\gamma _k$ and $\protect
  \mathaccentV {bar}016\gamma _q$ are obtained by inserting the expressions
  found for $\protect \mathaccentV {tilde}07Ek$ and $\protect \mathaccentV
  {tilde}07Eq$ into Eqs.~(\ref {eq:gK}) and (\ref {eq:BgQ}),
  respectively.}\BibitemShut {Stop}%
\bibitem [{\citenamefont {Volkov}\ and\ \citenamefont
  {Pankratov}(1985)}]{vol85}%
  \BibitemOpen
  \bibfield  {author} {\bibinfo {author} {\bibfnamefont {B.~A.}\ \bibnamefont
  {Volkov}}\ and\ \bibinfo {author} {\bibfnamefont {O.~A.}\ \bibnamefont
  {Pankratov}},\ }\bibfield  {title} {\bibinfo {title} {Two-dimensional
  massless electrons in an inverted contact},\ }\href@noop {} {\bibfield
  {journal} {\bibinfo  {journal} {Pis'ma Zh. Eksp. Teor. Fiz.}\ }\textbf
  {\bibinfo {volume} {42}},\ \bibinfo {pages} {145} (\bibinfo {year} {1985})},\
  \bibinfo {note} {[JETP Lett. \textbf{42}, 178 (1985)]}\BibitemShut {NoStop}%
\bibitem [{Note4()}]{Note4}%
  \BibitemOpen
  \bibinfo {note} {A different type of intrinsic semiconductor-nanoparticle
  sub-gap state discussed in Ref.~\cite {ser99} has its origin in a sign change
  of effective masses~\cite {lin85} instead of a band inversion.}\BibitemShut
  {Stop}%
\bibitem [{Note5()}]{Note5}%
  \BibitemOpen
  \bibinfo {note} {Using identity 10.2.2 from Ref.~\cite {abr64}, it can be
  shown that the spinors in Eqs.~(\ref {eq:evanAns}) are obtained from those in
  (\ref {eq:perpD}) by making the replacements $k = i\protect \mathaccentV
  {bar}016k$ and $\gamma _k = i\protect \tmspace +\thinmuskip {.1667em}
  \protect \mathaccentV {bar}016\gamma _{\protect \mathaccentV
  {bar}016k}$.}\BibitemShut {Stop}%
\bibitem [{\citenamefont {Haug}\ and\ \citenamefont {Koch}(2004)}]{hau04}%
  \BibitemOpen
  \bibfield  {author} {\bibinfo {author} {\bibfnamefont {H.}~\bibnamefont
  {Haug}}\ and\ \bibinfo {author} {\bibfnamefont {S.~W.}\ \bibnamefont
  {Koch}},\ }\href@noop {} {\emph {\bibinfo {title} {Quantum Theory of the
  Optical and Electronic Properties of Semiconductors}}},\ \bibinfo {edition}
  {4th}\ ed.\ (\bibinfo  {publisher} {World Scientific},\ \bibinfo {address}
  {Singapore},\ \bibinfo {year} {2004})\BibitemShut {NoStop}%
\bibitem [{\citenamefont {Pelant}\ and\ \citenamefont {Valenta}(2012)}]{pel12}%
  \BibitemOpen
  \bibfield  {author} {\bibinfo {author} {\bibfnamefont {I.}~\bibnamefont
  {Pelant}}\ and\ \bibinfo {author} {\bibfnamefont {J.}~\bibnamefont
  {Valenta}},\ }\href
  {https://doi.org/10.1093/acprof:oso/9780199588336.001.0001} {\emph {\bibinfo
  {title} {Luminescence Spectroscopy of Semiconductors}}}\ (\bibinfo
  {publisher} {Oxford University Press},\ \bibinfo {address} {Oxford, UK},\
  \bibinfo {year} {2012})\BibitemShut {NoStop}%
\bibitem [{\citenamefont {Gustafsson}\ \emph {et~al.}(1998)\citenamefont
  {Gustafsson}, \citenamefont {Pistol}, \citenamefont {Montelius},\ and\
  \citenamefont {Samuelson}}]{gus98}%
  \BibitemOpen
  \bibfield  {author} {\bibinfo {author} {\bibfnamefont {A.}~\bibnamefont
  {Gustafsson}}, \bibinfo {author} {\bibfnamefont {M.-E.}\ \bibnamefont
  {Pistol}}, \bibinfo {author} {\bibfnamefont {L.}~\bibnamefont {Montelius}},\
  and\ \bibinfo {author} {\bibfnamefont {L.}~\bibnamefont {Samuelson}},\
  }\bibfield  {title} {\bibinfo {title} {Local probe techniques for
  luminescence studies of low-dimensional semiconductor structures},\ }\href
  {https://doi.org/10.1063/1.368613} {\bibfield  {journal} {\bibinfo  {journal}
  {J. Appl. Phys.}\ }\textbf {\bibinfo {volume} {84}},\ \bibinfo {pages} {1715}
  (\bibinfo {year} {1998})}\BibitemShut {NoStop}%
\bibitem [{\citenamefont {Fomin}\ \emph {et~al.}(1998)\citenamefont {Fomin},
  \citenamefont {Gladilin}, \citenamefont {Devreese}, \citenamefont
  {Pokatilov}, \citenamefont {Balaban},\ and\ \citenamefont {Klimin}}]{fom98}%
  \BibitemOpen
  \bibfield  {author} {\bibinfo {author} {\bibfnamefont {V.~M.}\ \bibnamefont
  {Fomin}}, \bibinfo {author} {\bibfnamefont {V.~N.}\ \bibnamefont {Gladilin}},
  \bibinfo {author} {\bibfnamefont {J.~T.}\ \bibnamefont {Devreese}}, \bibinfo
  {author} {\bibfnamefont {E.~P.}\ \bibnamefont {Pokatilov}}, \bibinfo {author}
  {\bibfnamefont {S.~N.}\ \bibnamefont {Balaban}},\ and\ \bibinfo {author}
  {\bibfnamefont {S.~N.}\ \bibnamefont {Klimin}},\ }\bibfield  {title}
  {\bibinfo {title} {Photoluminescence of spherical quantum dots},\ }\href
  {https://doi.org/10.1103/PhysRevB.57.2415} {\bibfield  {journal} {\bibinfo
  {journal} {Phys. Rev. B}\ }\textbf {\bibinfo {volume} {57}},\ \bibinfo
  {pages} {2415} (\bibinfo {year} {1998})}\BibitemShut {NoStop}%
\bibitem [{Note6()}]{Note6}%
  \BibitemOpen
  \bibinfo {note} {In our notation, initial (final) states are labelled by
  quantum numbers $n, j, m, \kappa $ ($n', j', m', \kappa '$). Equation~(\ref
  {eq:dipole}) is a generalization of Eq.~(5.34) from Ref.~\cite {hau04} and,
  as the latter, relies on the assumption that the envelope-function part of
  electronic wave functions varies on much larger length scales than the basis
  functions.}\BibitemShut {Stop}%
\bibitem [{\citenamefont {Efros}\ and\ \citenamefont {Efros}(1982)}]{efr82}%
  \BibitemOpen
  \bibfield  {author} {\bibinfo {author} {\bibfnamefont {A.~L.}\ \bibnamefont
  {Efros}}\ and\ \bibinfo {author} {\bibfnamefont {A.~L.}\ \bibnamefont
  {Efros}},\ }\bibfield  {title} {\bibinfo {title} {Interband absorption of
  light in a semiconductor sphere},\ }\href@noop {} {\bibfield  {journal}
  {\bibinfo  {journal} {Fiz. Tekh. Poluprovodn.}\ }\textbf {\bibinfo {volume}
  {16}},\ \bibinfo {pages} {1209} (\bibinfo {year} {1982})},\ \bibinfo {note}
  {[Sov. Phys. Semicond. \textbf{16}, 772 (1982)]}\BibitemShut {NoStop}%
\bibitem [{\citenamefont {Manousakis}\ \emph {et~al.}(2017)\citenamefont
  {Manousakis}, \citenamefont {Altland}, \citenamefont {Bagrets}, \citenamefont
  {Egger},\ and\ \citenamefont {Ando}}]{man17}%
  \BibitemOpen
  \bibfield  {author} {\bibinfo {author} {\bibfnamefont {J.}~\bibnamefont
  {Manousakis}}, \bibinfo {author} {\bibfnamefont {A.}~\bibnamefont {Altland}},
  \bibinfo {author} {\bibfnamefont {D.}~\bibnamefont {Bagrets}}, \bibinfo
  {author} {\bibfnamefont {R.}~\bibnamefont {Egger}},\ and\ \bibinfo {author}
  {\bibfnamefont {Y.}~\bibnamefont {Ando}},\ }\bibfield  {title} {\bibinfo
  {title} {Majorana qubits in a topological insulator nanoribbon
  architecture},\ }\href {https://doi.org/10.1103/PhysRevB.95.165424}
  {\bibfield  {journal} {\bibinfo  {journal} {Phys. Rev. B}\ }\textbf {\bibinfo
  {volume} {95}},\ \bibinfo {pages} {165424} (\bibinfo {year}
  {2017})}\BibitemShut {NoStop}%
\bibitem [{\citenamefont {Bansil}\ \emph {et~al.}(2016)\citenamefont {Bansil},
  \citenamefont {Lin},\ and\ \citenamefont {Das}}]{ban16}%
  \BibitemOpen
  \bibfield  {author} {\bibinfo {author} {\bibfnamefont {A.}~\bibnamefont
  {Bansil}}, \bibinfo {author} {\bibfnamefont {H.}~\bibnamefont {Lin}},\ and\
  \bibinfo {author} {\bibfnamefont {T.}~\bibnamefont {Das}},\ }\bibfield
  {title} {\bibinfo {title} {Colloquium: Topological band theory},\ }\href
  {https://doi.org/10.1103/RevModPhys.88.021004} {\bibfield  {journal}
  {\bibinfo  {journal} {Rev. Mod. Phys.}\ }\textbf {\bibinfo {volume} {88}},\
  \bibinfo {pages} {021004} (\bibinfo {year} {2016})}\BibitemShut {NoStop}%
\bibitem [{\citenamefont {Herath}\ \emph {et~al.}(2014)\citenamefont {Herath},
  \citenamefont {Hewageegana},\ and\ \citenamefont {Apalkov}}]{her14}%
  \BibitemOpen
  \bibfield  {author} {\bibinfo {author} {\bibfnamefont {T.~M.}\ \bibnamefont
  {Herath}}, \bibinfo {author} {\bibfnamefont {P.}~\bibnamefont
  {Hewageegana}},\ and\ \bibinfo {author} {\bibfnamefont {V.}~\bibnamefont
  {Apalkov}},\ }\bibfield  {title} {\bibinfo {title} {A quantum dot in
  topological insulator nanofilm},\ }\href
  {https://doi.org/10.1088/0953-8984/26/11/115302} {\bibfield  {journal}
  {\bibinfo  {journal} {J. Phys.: Condens. Matter}\ }\textbf {\bibinfo {volume}
  {26}},\ \bibinfo {pages} {115302} (\bibinfo {year} {2014})}\BibitemShut
  {NoStop}%
\bibitem [{\citenamefont {Kundu}\ \emph {et~al.}(2011)\citenamefont {Kundu},
  \citenamefont {Zazunov}, \citenamefont {Yeyati}, \citenamefont {Martin},\
  and\ \citenamefont {Egger}}]{kun11}%
  \BibitemOpen
  \bibfield  {author} {\bibinfo {author} {\bibfnamefont {A.}~\bibnamefont
  {Kundu}}, \bibinfo {author} {\bibfnamefont {A.}~\bibnamefont {Zazunov}},
  \bibinfo {author} {\bibfnamefont {A.~L.}\ \bibnamefont {Yeyati}}, \bibinfo
  {author} {\bibfnamefont {T.}~\bibnamefont {Martin}},\ and\ \bibinfo {author}
  {\bibfnamefont {R.}~\bibnamefont {Egger}},\ }\bibfield  {title} {\bibinfo
  {title} {Energy spectrum and broken spin-surface locking in topological
  insulator quantum dots},\ }\href {https://doi.org/10.1103/PhysRevB.83.125429}
  {\bibfield  {journal} {\bibinfo  {journal} {Phys. Rev. B}\ }\textbf {\bibinfo
  {volume} {83}},\ \bibinfo {pages} {125429} (\bibinfo {year}
  {2011})}\BibitemShut {NoStop}%
\bibitem [{\citenamefont {Imura}\ \emph {et~al.}(2011)\citenamefont {Imura},
  \citenamefont {Takane},\ and\ \citenamefont {Tanaka}}]{imu11}%
  \BibitemOpen
  \bibfield  {author} {\bibinfo {author} {\bibfnamefont {K.-I.}\ \bibnamefont
  {Imura}}, \bibinfo {author} {\bibfnamefont {Y.}~\bibnamefont {Takane}},\ and\
  \bibinfo {author} {\bibfnamefont {A.}~\bibnamefont {Tanaka}},\ }\bibfield
  {title} {\bibinfo {title} {Spin {B}erry phase in anisotropic topological
  insulators},\ }\href {https://doi.org/10.1103/PhysRevB.84.195406} {\bibfield
  {journal} {\bibinfo  {journal} {Phys. Rev. B}\ }\textbf {\bibinfo {volume}
  {84}},\ \bibinfo {pages} {195406} (\bibinfo {year} {2011})}\BibitemShut
  {NoStop}%
\bibitem [{\citenamefont {Iorio}\ \emph {et~al.}(2016)\citenamefont {Iorio},
  \citenamefont {Perroni},\ and\ \citenamefont {Cataudella}}]{ior16}%
  \BibitemOpen
  \bibfield  {author} {\bibinfo {author} {\bibfnamefont {P.}~\bibnamefont
  {Iorio}}, \bibinfo {author} {\bibfnamefont {C.~A.}\ \bibnamefont {Perroni}},\
  and\ \bibinfo {author} {\bibfnamefont {V.}~\bibnamefont {Cataudella}},\
  }\bibfield  {title} {\bibinfo {title} {Quantum interference effects in
  {B}i$_2${S}e$_3$ topological insulator nanowires with variable cross-section
  lengths},\ }\href {https://doi.org/10.1140/epjb/e2016-70041-7} {\bibfield
  {journal} {\bibinfo  {journal} {Eur. Phys. J. B}\ }\textbf {\bibinfo {volume}
  {89}},\ \bibinfo {pages} {97} (\bibinfo {year} {2016})}\BibitemShut {NoStop}%
\bibitem [{\citenamefont {Virk}\ \emph {et~al.}(2018)\citenamefont {Virk},
  \citenamefont {Aut\`es},\ and\ \citenamefont {Yazyev}}]{vir18}%
  \BibitemOpen
  \bibfield  {author} {\bibinfo {author} {\bibfnamefont {N.}~\bibnamefont
  {Virk}}, \bibinfo {author} {\bibfnamefont {G.}~\bibnamefont {Aut\`es}},\ and\
  \bibinfo {author} {\bibfnamefont {O.~V.}\ \bibnamefont {Yazyev}},\ }\bibfield
   {title} {\bibinfo {title} {Electronic properties of one-dimensional
  nanostructures of the {${\mathrm{Bi}}_{2}{\mathrm{Se}}_{3}$} topological
  insulator},\ }\href {https://doi.org/10.1103/PhysRevB.97.165411} {\bibfield
  {journal} {\bibinfo  {journal} {Phys. Rev. B}\ }\textbf {\bibinfo {volume}
  {97}},\ \bibinfo {pages} {165411} (\bibinfo {year} {2018})}\BibitemShut
  {NoStop}%
\bibitem [{\citenamefont {Kernreiter}\ \emph {et~al.}(2013)\citenamefont
  {Kernreiter}, \citenamefont {Governale}, \citenamefont {Winkler},\ and\
  \citenamefont {Z\"ulicke}}]{ker13}%
  \BibitemOpen
  \bibfield  {author} {\bibinfo {author} {\bibfnamefont {T.}~\bibnamefont
  {Kernreiter}}, \bibinfo {author} {\bibfnamefont {M.}~\bibnamefont
  {Governale}}, \bibinfo {author} {\bibfnamefont {R.}~\bibnamefont {Winkler}},\
  and\ \bibinfo {author} {\bibfnamefont {U.}~\bibnamefont {Z\"ulicke}},\
  }\bibfield  {title} {\bibinfo {title} {Suppression of {C}oulomb exchange
  energy in quasi-two-dimensional hole systems},\ }\href
  {https://doi.org/10.1103/PhysRevB.88.125309} {\bibfield  {journal} {\bibinfo
  {journal} {Phys. Rev. B}\ }\textbf {\bibinfo {volume} {88}},\ \bibinfo
  {pages} {125309} (\bibinfo {year} {2013})}\BibitemShut {NoStop}%
\bibitem [{\citenamefont {Sabio}\ \emph {et~al.}(2010)\citenamefont {Sabio},
  \citenamefont {Sols},\ and\ \citenamefont {Guinea}}]{sab10}%
  \BibitemOpen
  \bibfield  {author} {\bibinfo {author} {\bibfnamefont {J.}~\bibnamefont
  {Sabio}}, \bibinfo {author} {\bibfnamefont {F.}~\bibnamefont {Sols}},\ and\
  \bibinfo {author} {\bibfnamefont {F.}~\bibnamefont {Guinea}},\ }\bibfield
  {title} {\bibinfo {title} {Two-body problem in graphene},\ }\href
  {https://doi.org/10.1103/PhysRevB.81.045428} {\bibfield  {journal} {\bibinfo
  {journal} {Phys. Rev. B}\ }\textbf {\bibinfo {volume} {81}},\ \bibinfo
  {pages} {045428} (\bibinfo {year} {2010})}\BibitemShut {NoStop}%
\bibitem [{\citenamefont {Berman}\ \emph {et~al.}(2013)\citenamefont {Berman},
  \citenamefont {Kezerashvili},\ and\ \citenamefont {Ziegler}}]{ber13}%
  \BibitemOpen
  \bibfield  {author} {\bibinfo {author} {\bibfnamefont {O.~L.}\ \bibnamefont
  {Berman}}, \bibinfo {author} {\bibfnamefont {R.~Y.}\ \bibnamefont
  {Kezerashvili}},\ and\ \bibinfo {author} {\bibfnamefont {K.}~\bibnamefont
  {Ziegler}},\ }\bibfield  {title} {\bibinfo {title} {Coupling of two {D}irac
  particles},\ }\href {https://doi.org/10.1103/PhysRevA.87.042513} {\bibfield
  {journal} {\bibinfo  {journal} {Phys. Rev. A}\ }\textbf {\bibinfo {volume}
  {87}},\ \bibinfo {pages} {042513} (\bibinfo {year} {2013})}\BibitemShut
  {NoStop}%
\bibitem [{\citenamefont {Trushin}\ \emph {et~al.}(2018)\citenamefont
  {Trushin}, \citenamefont {Goerbig},\ and\ \citenamefont {Belzig}}]{tru18}%
  \BibitemOpen
  \bibfield  {author} {\bibinfo {author} {\bibfnamefont {M.}~\bibnamefont
  {Trushin}}, \bibinfo {author} {\bibfnamefont {M.~O.}\ \bibnamefont
  {Goerbig}},\ and\ \bibinfo {author} {\bibfnamefont {W.}~\bibnamefont
  {Belzig}},\ }\bibfield  {title} {\bibinfo {title} {Model prediction of
  self-rotating excitons in two-dimensional transition-metal dichalcogenides},\
  }\href {https://doi.org/10.1103/PhysRevLett.120.187401} {\bibfield  {journal}
  {\bibinfo  {journal} {Phys. Rev. Lett.}\ }\textbf {\bibinfo {volume} {120}},\
  \bibinfo {pages} {187401} (\bibinfo {year} {2018})}\BibitemShut {NoStop}%
\bibitem [{\citenamefont {Sercel}\ \emph {et~al.}(1999)\citenamefont {Sercel},
  \citenamefont {Efros},\ and\ \citenamefont {Rosen}}]{ser99}%
  \BibitemOpen
  \bibfield  {author} {\bibinfo {author} {\bibfnamefont {P.~C.}\ \bibnamefont
  {Sercel}}, \bibinfo {author} {\bibfnamefont {A.~L.}\ \bibnamefont {Efros}},\
  and\ \bibinfo {author} {\bibfnamefont {M.}~\bibnamefont {Rosen}},\ }\bibfield
   {title} {\bibinfo {title} {Intrinsic gap states in semiconductor
  nanocrystals},\ }\href {https://doi.org/10.1103/PhysRevLett.83.2394}
  {\bibfield  {journal} {\bibinfo  {journal} {Phys. Rev. Lett.}\ }\textbf
  {\bibinfo {volume} {83}},\ \bibinfo {pages} {2394} (\bibinfo {year}
  {1999})}\BibitemShut {NoStop}%
\bibitem [{\citenamefont {Lin-Liu}\ and\ \citenamefont {Sham}(1985)}]{lin85}%
  \BibitemOpen
  \bibfield  {author} {\bibinfo {author} {\bibfnamefont {Y.~R.}\ \bibnamefont
  {Lin-Liu}}\ and\ \bibinfo {author} {\bibfnamefont {L.~J.}\ \bibnamefont
  {Sham}},\ }\bibfield  {title} {\bibinfo {title} {Interface states and
  subbands in {HgTe-CdTe} heterostructures},\ }\href
  {https://doi.org/10.1103/PhysRevB.32.5561} {\bibfield  {journal} {\bibinfo
  {journal} {Phys. Rev. B}\ }\textbf {\bibinfo {volume} {32}},\ \bibinfo
  {pages} {5561} (\bibinfo {year} {1985})}\BibitemShut {NoStop}%
\end{thebibliography}
%

\end{document}